\documentclass[fleqn,usenatbib]{mnras}

\usepackage{newtxtext,newtxmath}

\usepackage[T1]{fontenc}

\DeclareRobustCommand{\VAN}[3]{#2}
\let\VANthebibliography\thebibliography
\def\thebibliography{\DeclareRobustCommand{\VAN}[3]{##3}\VANthebibliography}

\usepackage{graphicx}	
\usepackage{amsmath}	
\usepackage{ulem}
\usepackage{multirow}

\newcommand{\V}{\mathcal{V}}
\newcommand{\cl}{$C_{\ell}(\Delta\nu)$}
\newcommand{\maps}{$C_{\ell}(\nu_a,\nu_b)$}

\newcommand{\HI}{\ion{H}{i}}
\newcommand{\obh}{$[\Omega_{\HI{}}b_{\HI{}}]$}

\defcitealias{Ch21}{Ch21} 
\defcitealias{P22}{Paper~I}

\title[Cross-polarization power spectrum]{Towards $21$-cm intensity mapping at $z=2.28$ with uGMRT using the tapered gridded estimator II: Cross-polarization power spectrum}

\author[A. Elahi et al.]{Kh. Md. Asif Elahi,$^{1}$\thanks{E-mail:asifelahi999@gmail.com}
Somnath Bharadwaj,$^{1}$\thanks{E-mail:somnath@phy.iitkgp.ac.in}
Abhik Ghosh,$^{2}$
Srijita Pal,$^{1}$
Sk. Saiyad Ali,$^{3}$
\newauthor
Samir Choudhuri,$^{4}$
Arnab Chakraborty,$^{5}$
Abhirup Datta,$^{6}$
Nirupam Roy,$^{7}$
Madhurima Choudhury,$^{6,8}$
\newauthor
Prasun Dutta$^{9}$
\\
\\
$^{1}$ Department of Physics and Centre for Theoretical Studies, IIT Kharagpur, Kharagpur 721 302, India\\
$^{2}$ Department of Physics, Banwarilal Bhalotia College, Asansol, West Bengal-713303, India\\
$^{3}$ Department of Physics, Jadavpur University, Kolkata 700032, India\\
$^{4}$ Centre for Strings, Gravitation and Cosmology, Department of Physics, Indian Institute of Technology Madras, Chennai 600036, India\\
$^{5}$ Department of Physics and McGill Space Institute, McGill University, Montreal, QC, Canada H3A 2T8\\
$^{6}$ Discipline of Astronomy, Astrophysics and Space Engineering, Indian Institute of Technology Indore, Indore 453552, India\\
$^{7}$ Department of Physics, Indian Institute of Science, Bangalore 560012, India\\
$^{8}$ ARCO (Astrophysics Research Center), Department of Natural Sciences, The Open University of Israel, 1 University Road, PO Box 808, Ra’anana 4353701, Israel\\
$^{9}$ Department of Physics, IIT (BHU), Varanasi, 221005 India
}

\date{Accepted XXX. Received YYY; in original form ZZZ}

\pubyear{2022}

\begin{document}
\label{firstpage}
\pagerange{\pageref{firstpage}--\pageref{lastpage}}
\maketitle

\begin{abstract}
    Neutral hydrogen (\HI{}) $21$-cm intensity mapping (IM) offers an efficient technique for mapping the large-scale structures in the universe. We introduce the `Cross' Tapered Gridded Estimator (Cross TGE), which cross-correlates two cross-polarizations (RR and LL) to estimate the multi-frequency angular power spectrum (MAPS) \cl{}. We expect this to mitigate several effects like noise bias, calibration errors etc., which affect the `Total' TGE which combines the two polarizations. Here we apply the Cross TGE on a $24.4 \,\rm{MHz}$ bandwidth uGMRT Band $3$ data centred at $432.8 \,\rm{MHz}$ aiming \HI{} IM at $z=2.28$. The measured \cl{} is modelled to yield maximum likelihood estimates of the foregrounds and the spherical power spectrum $P(k)$ in several $k$ bins.   Considering the mean squared brightness temperature fluctuations, we report a $2\sigma$ upper limit $\Delta_{UL}^{2}(k) \le (58.67)^{2} \, {\rm mK}^{2}$ at $k=0.804 \, {\rm Mpc}^{-1}$ which is a factor of $5.2$ improvement on our previous estimate based on the Total TGE. Assuming that the \HI{} traces the underlying matter distribution, we have modelled \cl{} to simultaneously estimate the foregrounds and $[\Omega_{\HI} b_{\HI}] $  where $\Omega_{\HI}$ and $b_{\HI}$ are the \HI{} density and linear bias parameters respectively. We obtain a best fit value of $[\Omega_{\HI}b_{\HI}]^2 = 7.51\times 10^{-4} \pm 1.47\times 10^{-3}$ which is consistent with noise. Although the $2\sigma$ upper limit $[\Omega_{\HI}b_{\HI}]_{UL} \leq 0.061$ is  $\sim 50$ times larger than the expected value, this is a considerable improvement over earlier works at this redshift.
\end{abstract}




\section{Introduction}

The $21$-cm line emission due to the hyperfine `spin flip' transition of the neutral hydrogen atom  (\HI{}) is a unique observational probe to the high redshift Universe. It can be used to study the large-scale structures in the Universe, probe the ionization state of the inter-galactic medium, constrain the Dark Energy equation of state, put independent limits on various cosmological parameters, and quantify non-Gaussianity \citep{BA5,w08,loeb08,mao2008,Bh09,Visbal_2009,morales10, bagla2010,Hazra2012, prichard12, ansari2012,battye2013,Bull15,long2022}. At post-Epoch of Reionization (post-EoR; $z<6$), the vast majority of \HI{} is resided in discrete high density clouds having column densities larger than $2\times10^{20}\,\,{\rm atoms}\,\,{\rm cm}^{-2}$ \citep{Wolfe95, ho2021}. Instead of resolving these faint, discrete objects individually, the \HI{} Intensity Mapping (IM) approach integrates the $21$-cm emission over the large volumes of observation \citep{BNS, BS01,bh_sri2004}, enabling a full three-dimensional map of the \HI{} distribution.

Several low-redshift $(z < 1)$ single dish experiments (e.g. \citealt{Pen2009a, chang10, masui2013, SW13, Anderson2018, Wolz2021}) have cross-correlated IM signal with optical galaxy redshift surveys (e.g. DEEP2; \citealt{Newman_2013}) to constrain the \HI{} distribution. Recently, cross-correlating with the eBOSS \citep{Dawson_2016} galaxy catalogues, the CHIME\footnote{\url{https://chime-experiment.ca/en/}} \citep{chimeIM} interferometer has detected the $21$-cm signal in the redshift range $0.78<z<1.43$ \citep{chime22}. However, an auto-correlation (i.e., not in cross-correlation with other probes) detection of the redshifted $21$-cm signal is yet to be made. One of the primary science goals of the ongoing and upcoming IM experiments, such as BINGO\footnote{\url{https://bingotelescope.org/}} \citep{Wuensche_2019}, HIRAX\footnote{\url{https://hirax.ukzn.ac.za/}} \citep{Newburgh16}, MeerKAT\footnote{\url{https://www.sarao.ac.za/science/meerkat/}} \citep{Kennedy21}, and the Tianlai project\footnote{\url{http://tianlai.bao.ac.cn/}} \citep{tian} is to measure  the Baryon Acoustic Oscillation (BAO) in the post-EoR $21$-cm power spectrum (PS). Along with BAO, the next-generation IM surveys with the SKA\footnote{\url{https://www.skatelescope.org/}} \citep{SKA15}, recently upgraded OWFA\footnote{\url{http://rac.ncra.tifr.res.in/ort.html}} \citep{OWFA} hold the promise to extract an ample amount of cosmological information through the $21$-cm PS. 

As a way forward in this direction, a few upper limits on the amplitude of the redshifted \HI{} signal have been placed using the Giant Metrewave Radio Telescope (GMRT\footnote{\url{http://www.gmrt.ncra.tifr.res.in/}}; \citealt{swarup91}). \cite{ghosh1, ghosh2} have used $610 \, {\rm MHz}$ $(z=1.32)$ GMRT data to constrain $[\Omega_{\HI} b_{\HI}] < 0.11$ at $3\sigma$ level, where $\Omega_{\HI}$ is the comoving \HI{} mass density in units of the present critical density and $b_{\HI}$ is the \HI{} bias parameter. More recently, using the upgraded GMRT (uGMRT; \citealt{uGMRT}) data, \cite{Ch21} (hereafter Ch21) have put multi-redshift constraints on the amplitude of $21$-cm PS $\Delta_{UL}^{2}(k) \leq (58.87)^2,\,(61.49)^2,\,(60.89)^2\,\rm{and}\,(105.85)^2 \,\rm{mK}^2$ at $k = 1\,\rm{Mpc}^{-1}$. These values translate to the upper limits of $[\Omega_{\HI} b_{\HI}]_{UL} \leq 0.09,\,0.11,\,0.12\,\rm{and}\,0.24$ at $z=1.96,\,2.19,\,2.62\,\rm{and}\,3.58$ respectively.

The biggest challenge to a high redshift $21$-cm IM experiment is perhaps the foregrounds which are $4-5$ orders of magnitude brighter than the predicted signal (e.g. \citealt{shaver99, dmat1, santos05, ali, ali14}). The diffuse galactic synchrotron emission (DGSE) from our Galaxy and the extragalactic point sources (EPS) -- which are the diffused emission from the external galaxies -- are the most dominant foreground components considering the post-EoR observations \citep{haslam81, haslam82, reich88, condon1989, cress1996, wilman2003, blake2004, owen2008, singal2010, condon2012, randall2012, zheng17}. The wide-field foregrounds (mainly the EPS) are very challenging to deal with even with the existing foreground removal \citep{jelic08, bowman09, paciga11, chapman12, trott1, trott16, mertens18} and `foreground avoidance'  \citep{adatta10, vedantham12, thyag13, pober13, pober14, liu14a, liu14b, dillon14, dillon15} techniques. 

The Multi-frequency Angular Power Spectrum (MAPS; \citealt{Zaldarriaga2004, santos05, KD07}) \maps{} which characterizes the second order statistics of the sky signal jointly as a function of the angular multipole $\ell$ and frequencies $\nu$, is a promising statistics to quantify the $21$-cm signal  \citep{Mondal2018,Mondal19} and  distinguish it from  the foregrounds \citep{liu12,Trott2022}. In the present work we have used the MAPS \cl{} which only depends on the frequency separation $\Delta\nu= \mid \nu_a - \nu_b \mid$. This is adequate  when the statistical properties of the 21-cm signal do not vary significantly across  the frequency bandwidth under consideration. Foregrounds and the $21$-cm signal are expected to show contrasting behaviour in \cl{}. Being spectrally smooth, Foregrounds are expected to show little or no variation with $\Delta\nu$ compared to the $21$-cm signal which is expected to decorrelate with increasing $\Delta\nu$ \citep{BS01,BA5, santos05, ali, ali14}. However, the wide-field point sources introduce oscillatory patterns along $\Delta\nu$ in the estimated $C_{\ell}(\Delta\nu)$ due to the inherent frequency response of the radio interferometers \citep{ghosh1,ghosh2}. These oscillations, whose frequency increases at larger $\ell$ due to baseline migration, also manifest themselves as the `foreground wedge' \citep{adatta10, Morales_2012} structure in the estimated cylindrical PS $P(k_{\perp},k_{\parallel})$ (\citealt{P22}; hereafter, \citetalias{P22}). The frequency structures in the $C_{\ell}(\Delta\nu)$, or the foreground wedge, jeopardise faithful foreground removal and recovery of the $21$-cm signal from the measured visibility data. Additionally, considering foreground avoidance, various instrumental systematics,  such as gain variations, primary beams, polarization leakage, calibration errors, missing channels flagged due to Radio Frequency Interference (RFI) etc., extend the foreground wedge to much higher $k_{\parallel}$ values, consequently reducing the otherwise foreground-free `$21$-cm window' (TW) \citep{bowman09, pober16, Thyagarajan_2016}. 

The Tapered Gridded Estimator (TGE; \citealt{samir14, samir16, samir17}) is a visibility-based $21$-cm PS estimator which allows us to taper the sky response to suppress the wide-field foreground contributions arising from the side-lobe or periphery of the primary beam pattern. Additionally, it reduces computational load by using gridded visibilities, and internally subtracts out the positive-definite noise bias to produce unbiased estimates of the measured quantities. The TGE has been used to characterise the angular power spectrum $C_{\ell}$ of the foregrounds at EoR frequencies \citep{samir17a, samir20} as well as post-EoR frequencies \citep{Cha1, Cha2, Aishrila2020}. \cite{Bh18} further developed upon this to introduce a MAPS-based TGE which first estimates the MAPS, and, from it, the PS, effectively dealing with the missing frequency channels in the visibility data while preserving all the qualities mentioned above. \cite{Pal20} have used the MAPS-based TGE to estimate the MAPS and PS of the redshifted \HI{} signal from EoR using an $8\,\,{\rm MHz}$ GMRT data set observed at $153\,\,{\rm MHz}$. 

In this work, we consider an observation of $25$ hours over $4$ nights from the ELAIS-N1 field using a $200$ MHz bandwidth at Band 3 $(300-500\,{\rm MHz})$ of uGMRT. This data was first introduced in \cite{Cha1}, and in a follow-up work, \cite{Cha2}  have presented the flagging, calibration, imaging, and point source subtraction from this data and also used the 2D TGE to study the angular and spectral variation of $C_{\ell}(\nu)$ for the DGSE. \citetalias{Ch21} have conducted a multi-redshift analysis of this data  using a  delay spectrum approach to estimate the PS of the 21-cm intensity mapping signal.  In this approach  the missing frequency channels (flagged due to RFI) introduce ringing artefacts in the delay space, which can cause additional foreground leakage and corrupt the estimated PS. The one-dimensional (1D) CLEAN \citep{Parsons_2009} and the Least Square Spectral Analysis (LSSA; \citealt{Trott2016}) are the two commonly used techniques which allow one to compensate for the missing frequency channels. Many recently developed algorithms, such as DAYENU filter \citep{Ewall-Wice2021}, Gaussian Process Regression (GPR; \citealt{mertens20, Kern2021, Trott2020}) and Gaussian Constrained Realizations (GCR; \citealt{Kennedy2022}), have also aimed for an accurate recovery of the $21$-cm PS from an RFI-contaminated data. \cite{Chakraborty_2022} have recently compared the 1D CLEAN and LSSA with simulated and actual visibility data to check which of the methods work better. 

The TGE is capable of recovering the $21$-cm signal even when $80\%$ data from randomly selected frequency channels are flagged \citep{Bh18}. The TGE first correlates the visibility data across frequency channels to estimate $C_{\ell}(\Delta \nu)$, and estimates the PS from \cl{}.  Even if there are a substantial number of  missing frequency channels in the visibility data, it is possible that there are no missing  frequency separations $\Delta \nu$ in the estimated  $C_{\ell}(\Delta \nu)$. The entire procedure uses only the available frequency channels to estimate the PS, it is not essential to make up for any missing frequency channels. In \citetalias{P22} we have used the TGE on a $24.4\,{\rm MHz}$ bandwidth data at $432.8\,{\rm MHz}$ $(z=2.28)$ from the same observation where $55\%$ of the data were flagged. \citetalias{P22} further used foreground avoidance approach to constrain the mean squared brightness temperature fluctuations of the redshifted \HI{} signal with a $2\sigma$ upper limit of $\Delta_{UL}^{2}(k) \le (133.97)^{2} \, {\rm mK}^{2}$ at $k=0.347 \, {\rm Mpc}^{-1}$ which corresponds to an upper limit $[\Omega_{\HI} b_{\HI}]_{UL} \le 0.23$ at $z=2.28$. The quoted upper limit was found to be $\sim7$ times larger than what \citetalias{Ch21} found at a close redshift of $z=2.19$ ($\nu_c = 445\,{\rm MHz}$). 

\begin{figure}
    \centering
	\includegraphics[width=0.85\columnwidth, bb=96. 120. 830. 389.]{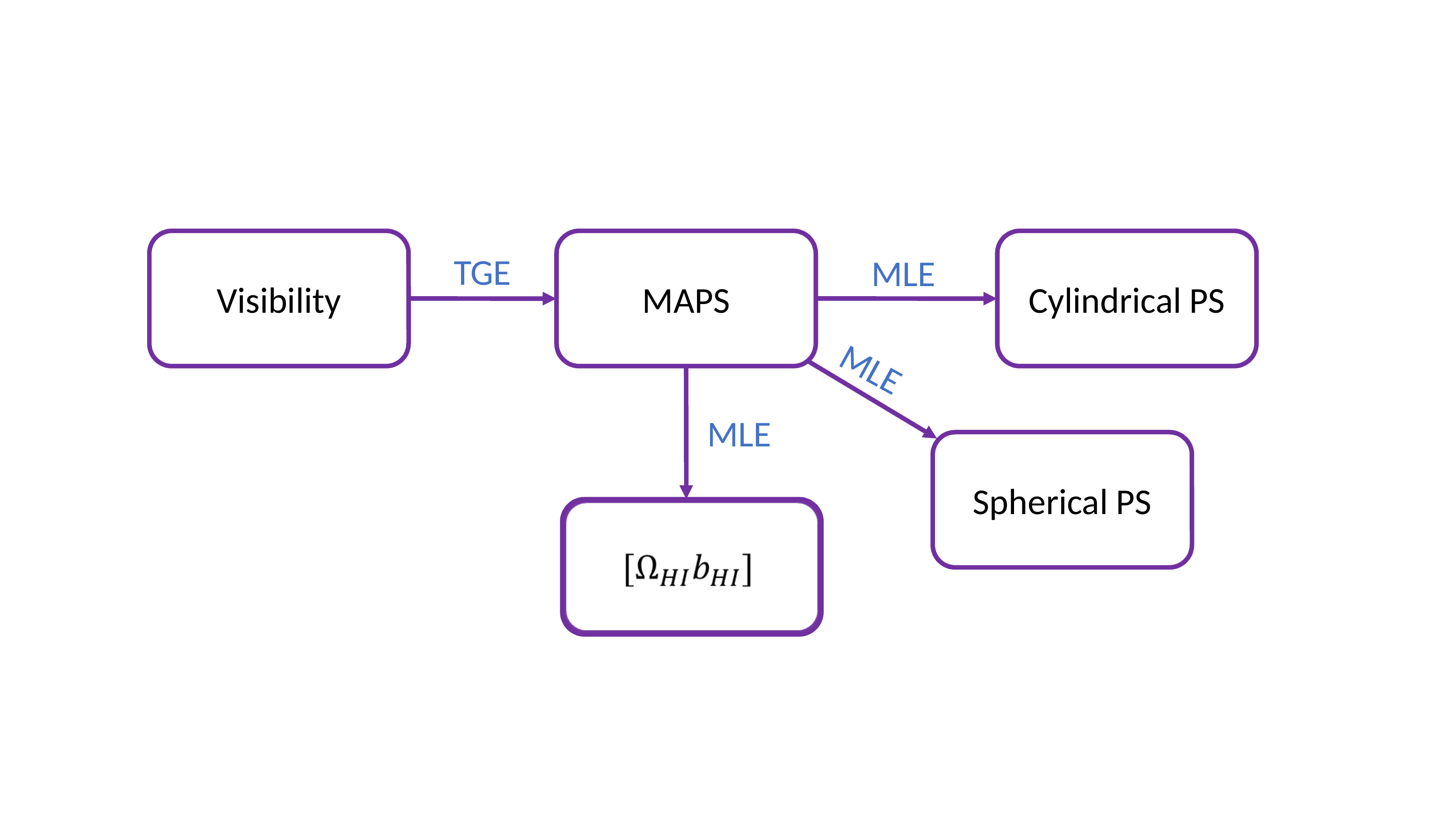}
    \caption{Flowchart of the paper, highlighting the main steps considered for the $21$-cm IM from the calibrated visibilities.}
    \label{fig:tge_diagram}
\end{figure}

The present work considers the same data as used in \citetalias{P22}, with two key differences introduced in the analysis technique. Firstly, we define the `Cross' TGE for MAPS which cross-correlates the two mutually orthogonal (Cross) polarization states (RR and LL) of the visibilities. We expect this cross-correlation approach to mitigate a number of issues, such as noise bias, calibration errors etc., which affect the `Total' TGE (used in \citetalias{P22}) where the two polarizations are combined. Further, it is expected that this approach will also reduce contributions from polarization-dependent foregrounds and systematics. Secondly, we have introduced a novel Maximum Likelihood Estimator (MLE) which  estimates the spherical PS $P(k)$ of the $21$-cm signal directly from the estimated $C_{\ell}(\Delta\nu)$ without explicitly referring to the cylindrical PS $P(k_{\perp},k_{\parallel})$. The MLE we present utilizes the statistical isotropy of the $21$-cm signal that differentiates it from the foregrounds. The MLE is expected to be robust to outliers \citep{Huber}, and is optimal as we use inverse noise covariance weightage in the likelihood. Apart from estimating $P(k)$, we have also used the MLE on the full data set (or a subset) to constrain the single parameter $[\Omega_{\HI} b_{\HI}]$, thus maximizing the signal-to-noise ratio. A simplified flowchart of our present work is presented in Figure~\ref{fig:tge_diagram}.

We have arranged the paper in the following way. We first summarize the observations and preliminary processing of the data in Section~\ref{sec:data}. Next, we present the formalism for the TGE along with the estimated MAPS in Section~\ref{sec:maps}, and the formalism for cylindrical power spectrum estimation along with the estimated $P(k_{\perp},k_{\parallel})$ in Section~\ref{sec:cylps}. In Sections~\ref{sec:BMLE} and ~\ref{sec:omb} we present the MLE for obtaining  the spherical PS $P(k)$ and $[\Omega_{\HI} b_{\HI}]$ respectively, and also present the corresponding results. We have summarized our findings in Section~\ref{sec:conclusion}.

Same as \citetalias{P22}, we have used a $\Lambda \rm{CDM}$ cosmology with $\Omega_{m} = 0.309$,  $h = 0.67$, $n_s = 0.965$, and $\Omega_{b}h^2 = 0.0224$ \citep{Planck18f}.

\section{Data Description}
\label{sec:data}
We have observed the ELAIS-N1 field during May 2017 for 25 hours over four nights using a $200$ MHz bandwidth at the Band 3 $(300 - 500 \,{\rm MHz})$  of uGMRT with a frequency resolution  $(\Delta \nu_c)$ of $24.4\,\rm{kHz}$ and an integration time of $2\rm{s}$. The detailed description of the data, along with flagging, calibration, imaging and  point source subtraction, are presented in \cite{Cha2}. We have used the resulting flagged, calibrated, point source subtracted visibility data for the entire analysis presented here. Note that polarization calibration is not performed on this data. 

The subset of the above data  which has been analysed here is the same as that in \citetalias{P22}, with the difference that we have restricted the  baselines to a smaller range $\mid \mathbf{U}\mid<1000\lambda$ where the baseline coverage is found  to be denser and nearly uniform (Figure 1 of \citetalias{P22}). The data covers a $24.4$ MHz bandwidth with a central frequency $\nu_c = 432.84 \, \rm{MHz}$. 

Considering the visibility data  which we have analyzed here, $\V{}_i^{x}(\nu_a)$  refers to a visibility measured at the baseline $\mathbf{U}_i$, frequency $\nu_a$ and polarization $x$. The present data contains two circularly polarized states RR and LL.

\section{The TGE for MAPS}
\label{sec:maps}

The multi-frequency angular power spectrum (MAPS) $C_{\ell}(\nu_a, \nu_b)$  quantifies  the statistical properties of the sky signal jointly as a function of the angular multipoles and frequencies. The brightness temperature fluctuations in the sky are decomposed in terms of  spherical harmonics $Y_{\ell}^{\rm m}(\hat{\mathbf{n}})$ as
\begin{equation}
    \delta T_{\rm b} (\hat{\mathbf{n}},\,\nu)=\sum_{\ell,m} a_{\ell {\rm m}} (\nu) \,
    Y_{\ell}^{\rm m}(\hat{\mathbf{n}}) \,.
    \label{eq:alm}
\end{equation}
We use this to define the  MAPS  as \citep{Zaldarriaga2004, santos05, KD07}
\begin{equation}
    C_{\ell}(\nu_a, \nu_b) = \big\langle a_{\ell {\rm m}} (\nu_a)\, a^*_{\ell
      {\rm m}} (\nu_b) \big\rangle\,.
    \label{eq:cl}
\end{equation}
Here $\langle  ... \rangle$ denotes an ensemble average  over different statistically independent realizations of the random  field  $\delta T_{\rm b} (\hat{\mathbf{n}},\,\nu)$. 

The Tapered Gridded Estimator (TGE) uses the measured visibilities to estimate MAPS. We note that the present analysis does not incorporate  baseline migration. Considering a fixed antenna pair,   the baseline $\mathbf{U}$  is held fixed at the value corresponding to the central frequency $\nu_c$.  The details of the visibility based TGE are given in \citetalias{P22} (also  \citealt{Bh18} and \citealt{Pal20}). Here we briefly summarize the salient features of the formalism, and extend it to consider the polarization. We introduce a rectangular grid in the $uv$-plane and  calculate $\V_{cg}^{x}(\nu_a)$ the convolved-gridded visibility for every grid point $\mathbf{U}_g$ using 
\begin{equation}
    \V_{cg}^{x}(\nu_a) = \sum_i  \tilde{w}(\mathbf{U}_g-\mathbf{U}_i) \, \V{}_i^{x}(\nu_a) \, F_i^{x}(\nu_a) \,.
	\label{eq:vcgx}
\end{equation}
Here  $F_i^{x}(\nu_a)$ is $0$ if the  visibility is flagged and $1$ otherwise, and   $\tilde{w}(\mathbf{U})$ is the Fourier transform of a  suitably chosen window function ${\mathcal W}(\theta)$ which is introduced to taper the primary beam (PB) of the telescope far away from the phase center.

The main lobe of the PB of any telescope with a circular aperture can be approximated as $\mathcal{A}(\theta)=e^{-\theta^{2}/\theta^{2}_{0}}$,  where $\theta_{0}\sim 0.6 \times \theta_{\rm FWHM}$, $\theta_{\rm FWHM}$ is the full width at half maxima of $\mathcal{A}(\theta)$\citep{BS01, samir14}. Here we have used a Gaussian window function ${\mathcal W}(\theta)=e^{-\theta^{2}/[f \theta_{0}]^2}$  where  the tapering parameter `$f$' controls the degree to which the PB pattern is tapered.  Here   $f>1$  provides minimal tapering,  and $f<1$  highly suppresses the sky response  away from the phase center. We had considered different values of $f$ in the range $0.6 \le f \le 5$  in \citetalias{P22} where we had found that it is possible to reduce  oscillations (along frequency) in MAPS by reducing the value of $f$. However, this improvement was found to saturate around $f=0.6$ which  provides the best results. Based on this, we have used   $f=0.6$  for the entire analysis presented here. 

Here we assume that the $21$-cm signal is unpolarized, and we treat the two polarizations (RR and LL) as independent measurements of the same $21$-cm  signal. 
In \citetalias{P22} we have combined the two polarizations using 
\begin{equation}
    \V_{cg}(\nu_a) = \V_{cg}^{RR}(\nu_a) + \V_{cg}^{LL}(\nu_a) \,.
    \label{eq:vcg}
\end{equation}
and used this to define the TGE for MAPS
\begin{align}
   \hat{E}_g & (\nu_a,\,\nu_b) = M_g^{-1}(\nu_a,\nu_b) {\mathcal Re} \Big[\V_{cg}(\nu_a)   \V_{cg}^{*}(\nu_b) \nonumber \\ 
    & - \sum_{x,\, i} F_i^{x}(\nu_a)F_i^{x}(\nu_b)  \mid \tilde{w}(\mathbf{U}_g-\mathbf{U}_i) \mid^2   \V_i^{x}(\nu_a) \V_i^{*x}(\nu_b)  \Big]  
    \label{eq:selfmaps}
\end{align}
where ${\mathcal Re}[..]$ implies the real part of the expression within the brackets $[..]$ and  $M_g(\nu_a,\nu_b)$ is a normalization factor. In principle, it is adequate to consider the correlation    $\V_{cg}(\nu_a)   \V_{cg}^{*}(\nu_b)$  in order to estimate 
$C_{\ell}(\nu_a, \nu_b)$, except for the fact that we have an additive  noise bias when  $\nu_a=\nu_b$. The second term in the square brackets, which subtracts out the  correlation of a visibility with itself,  is introduced to remove the noise bias.

Instead of  combining the two polarizations (equation~\ref{eq:vcg}), in the present  work  we have used the  correlation of the two cross-polarizations  $\V_{cg}^{RR}$ and $\V_{cg}^{LL}$ to estimate $C_{\ell}(\nu_a, \nu_b)$. The cross-polarization correlation TGE for MAPS is defined as
\begin{align}
   \hat{E}_g (\nu_a,\,\nu_b) = M_g^{-1}(\nu_a,\nu_b) {\mathcal Re} \Big[  &  \V_{cg}^{RR}(\nu_a)  \V_{cg}^{*LL}(\nu_b)  \nonumber \\ 
    & + \V_{cg}^{LL}(\nu_a) \V_{cg}^{*RR}(\nu_b)   \Big]  \,. 
    \label{eq:crossmaps}
\end{align}
Since the noise in the two polarizations are uncorrelated, 
equation~(\ref{eq:crossmaps}) has the advantage that it is not necessary to account for any  noise bias in the cross-polarization estimator.  We may also expect some further advantages if the calibration errors, foregrounds and other systematics in the two polarizations are partially uncorrelated. We note that the estimator in  equation~(\ref{eq:selfmaps}) contains both the self-polarization correlations (RR $\times$ RR and LL $\times$ LL) and the cross-polarization correlations (RR $\times$ LL). In the subsequent discussion we refer to the TGE in equations (\ref{eq:selfmaps}) and  (\ref{eq:crossmaps}) as `Total' and `Cross' respectively. We have validated (Appendix~\ref{sec:validation}) the  Cross TGE  using simulations  which  incorporate the same flagging, frequency  and  baseline coverage of the actual data. The validation of the Total estimator is given in \citetalias{P22}.

We now discuss how we have determined the normalization factor  $M_g^{-1}(\nu_a,\nu_b)$ for the Cross estimator. We first simulate multiple realizations of $[\delta T_{\rm b} (\hat{\mathbf{n}},\,\nu)]_{\rm uMAPS}$  the sky signal corresponding to a Gaussian random field having a unit MAPS (uMAPS; $C_{\ell}(\nu_a, \nu_b)=1$). We use this  sky signal to simulate the corresponding visibilities $ [\V_i^{x}(\nu_a)]_{\rm uMAPS}$ at the baselines, frequency channels and polarizations identical to the data. These simulations incorporate both  baseline migration and the  frequency dependence of the  telescope's PB.  The flagging of the actual data $F_i^{x}(\nu_a)$ has been applied to the simulated visibilities  $ [\V_i^{x}(\nu_a)]_{\rm uMAPS}$ and  used to obtain 
\begin{align}
    {M}_g(\nu_a,\nu_b) = {\mathcal Re} \Big[\V_{cg}^{RR} & (\nu_a)  \V_{cg}^{*LL}(\nu_b)  \nonumber \\ 
    & + \V_{cg}^{LL}(\nu_a) \V_{cg}^{*RR}(\nu_b)   \Big]_{\rm {uMAPS}} \,.  
    \label{eq:uMAPS}
\end{align}
We have averaged over multiple realizations of the simulated uMAPS to reduce the statistical uncertainties in the estimated $M_g$. For the subsequent analysis, we have used $50$ realizations of uMAPS to estimate $M_{g}$. Note that $M_{g}$ for the Total estimator (equation~\ref{eq:selfmaps}) is different 
from that given by equation~(\ref{eq:uMAPS}),  and the relevant equation for the Total estimator is presented in \citetalias{P22}.

Both the estimators (equations~\ref{eq:selfmaps} and \ref{eq:crossmaps}) give unbiased estimate of the MAPS {\it i.e.}  $\langle \hat{E}_g\rangle =C_{\ell_g}$  at the grid point ${\bf U}_g$ which corresponds to an angular multipole  $\ell_g=2\,\pi\,\mid {\bf U}_g \mid$. Incorporating the fact that the  statistics of the   $21$-cm signal is   isotropic on the plane of the sky,  we combine  the $\hat{E}_g$ at different grid points ${\bf U}_g$ within annular bins in the $uv$-plane.  The bin averaged TGE is defined as, 
\begin{equation}
    \hat{E}_a = \frac{\sum_g w_g  \hat{E}_g}{\sum_g w_g } 
\label{eq:binnedtge}
\end{equation}
where the sum is over all the grid points ${\bf U}_g$ in the $a^{\rm{th}}$ $\ell$ bin   and  the $w_g$'s are the corresponding weights. Here, we have used $w_g=M_g$  which implies that the weight is proportional to the baseline density in that particular grid point. The ensemble average of $\hat{E}_a$ gives an unbiased estimate of the bin averaged MAPS $\bar{C}_{\bar{\ell_a}}$ at the bin averaged angular multipole $\bar{\ell_a} = \frac{\sum_g w_g \ell_g}{\sum_g w_g}$. We subsequently use $C_{\ell_a}$ and $\ell_a$ to denote the bin-averaged values $\bar{C}_{\bar{\ell_a}}$ and $\bar{\ell_a}$ respectively. 

The post-reionization $21$-cm signal evolves relatively gradually with $z$ (e.g. \citealt{Deb16}), and it is quite reasonable to assume this to be ergodic (statistically homogeneous) along the line-of-sight direction for the $24.4 \, \rm{MHz}$  bandwidth, which corresponds to the redshift interval $\Delta z = 0.19$ $(z = 2.19-2.38)$, considered here. Instead of considering the entire covariance  $C_{\ell}(\nu_a, \nu_b)$, it is now adequate to consider $C_{\ell}(\Delta \nu)$ which is  a function of the frequency separations $\Delta \nu = |\nu_a - \nu_b|$. However, we note that $C_{\ell}(\nu_a, \nu_b)$ is a more accurate statistics (see, e.g. \citealt{Mondal2018,Mondal2022}) for a wide-band and high-redshift data (such as \citealt{Trott2020}).

\begin{figure}
	\includegraphics[width=\columnwidth]{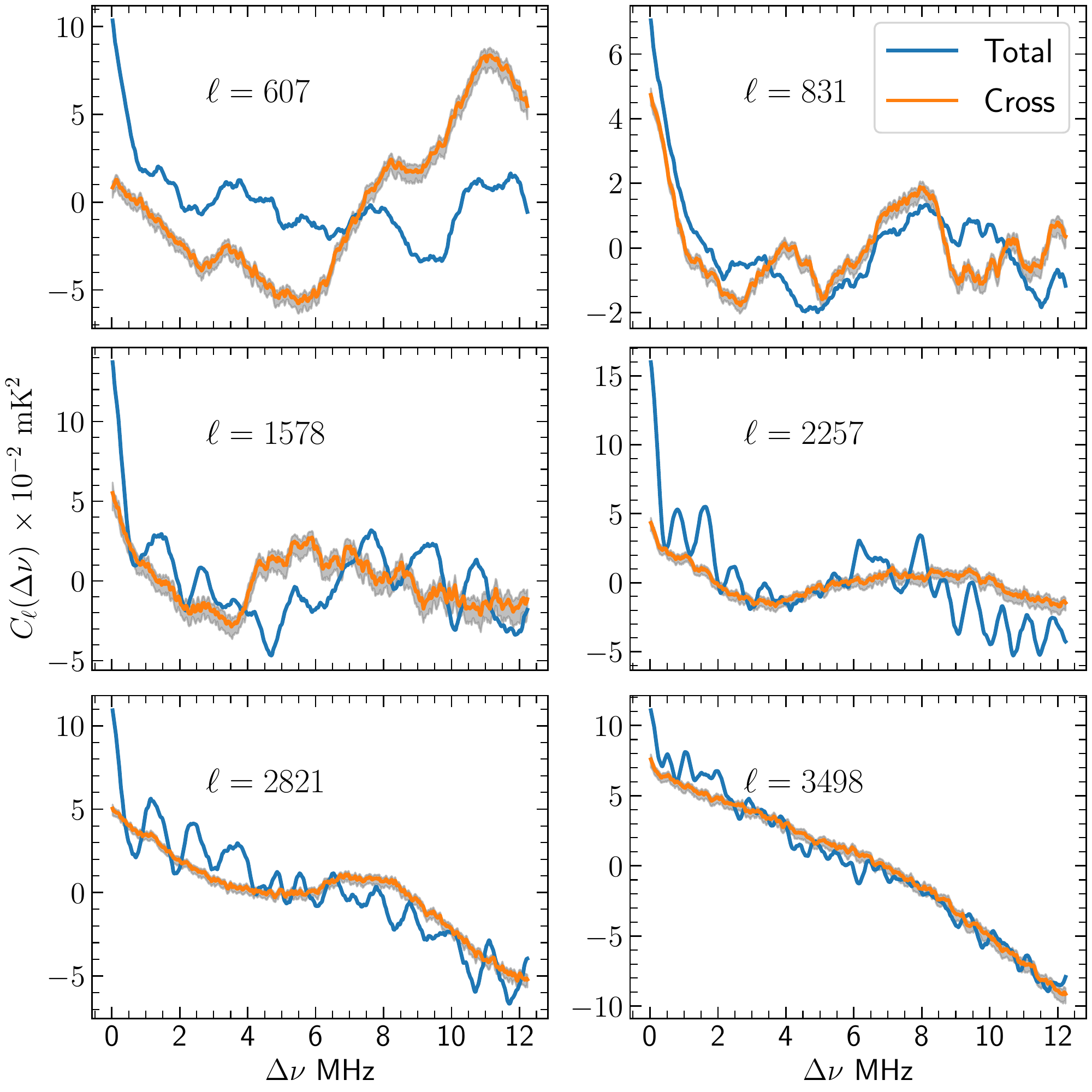}
    \caption{A comparison of mean-subtracted Total (blue) and Cross  (orange) MAPS \cl{}  for different $\ell$-values.   The  grey shaded regions show the $3\sigma$ error bars.}
    \label{fig:cldnu}
\end{figure}

We have divided the ${\bf U}$ range  $U \le 1000\lambda$  into $10$ bins of equal  linear spacing,  and  evaluated $C_{\ell}(\Delta \nu)$  using the Cross estimator  (equation~\ref{eq:crossmaps}).  The estimated $C_{\ell}(\Delta \nu)$ are rather noisy at  large $\Delta \nu$, and following  \citetalias{P22} we have only used  $\Delta \nu \le  12.2 \, \rm{MHz}$ for the subsequent analysis. The $\ell$ bins used here are somewhat  different from those in \citetalias{P22}, however  the  $\ell$ values  roughly match for the  first $6$ bins. Compared to \citetalias{P22} which has used the Total estimator, we find   that the  $C_{\ell}(\Delta \nu)$ values obtained here 
have a different vertical offset  which  corresponds to a  difference in the  $\Delta \nu$ independent  DC component of  $C_{\ell}(\Delta \nu)$. 
This DC component of $C_{\ell}(\Delta \nu)$ only affects the lowest line-of-sight mode $k_{\parallel} = 0$ in $P(k_{\perp}, k_{\parallel})$ the cylindrical PS. The $k_{\parallel} = 0$ mode is usually foreground dominated, and we do not use it to constrain the $21$-cm signal. Figure~\ref{fig:cldnu} shows the DC subtracted $C_{\ell}(\Delta \nu)$ obtained using both the Cross and  the  Total estimators for the first $6$   $\ell$ bins. The  grey shaded regions show the $3\sigma$ errors for  the Cross $C_{\ell}(\Delta \nu)$. These errors  were estimated using simulations as described in \citetalias{P22}, and also later in this paper. 

We expect the estimated $C_{\ell}(\Delta \nu)$ (both Cross and Total)  to be dominated by various  foreground components, mainly the  diffuse Galactic synchrotron emission (DGSE) and the radiation from unsubtracted extragalactic point sources (EPS). It has been reported in earlier studies that the measured $C_{\ell}$  is dominated by the DGSE at larger angular scales and by the residual point sources  at smaller angular scales \citep{bernardi09,ghosh3,samir17a, Cha2}. While both of these are expected to have intrinsically smooth frequency spectra, various observational effects introduce frequency-dependent structures in the estimated $C_{\ell}(\Delta \nu)$. For example, baseline migration, bandpass calibration errors and polarization leakage introduce  oscillations along $\Delta \nu$ in $C_{\ell}(\Delta \nu)$ . 

Considering the different panels of Figure~\ref{fig:cldnu}, we see that the two different estimates of $C_{\ell}(\Delta \nu)$  have comparable values once the DC is subtracted out. Considering the  $\Delta \nu$ dependence, for many of the $\ell$ bins  we find very similar slowly varying patterns in both the estimates. The degree of similarity appears to increase as we go to the larger $\ell$ bins. The Total  $C_{\ell}(\Delta \nu)$ de-correlates very sharply when $\Delta \nu$ is increased from $0$ to $1 \, {\rm MHz}$, and it also exhibits rapid oscillations at larger $\Delta \nu$. These rapid variations are considerably diminished in the Cross $C_{\ell}(\Delta \nu)$ which exhibits a much smoother $\Delta \nu$ dependence. 

In the lowest $\ell$ bin, the  Cross and Total  $C_{\ell}(\Delta \nu)$ are found to differ in their  $ \Delta \nu$ dependence.  We expect  $C_{\ell}(\Delta \nu)$ in this bin to be DGSE dominated (\citealt{Cha2}). The differences between the two estimates of $C_{\ell}(\Delta \nu)$  may arise due  to  polarized  structure in the DGSE (\citealt{ulipen2009}). Further, these differences may also arise from differences in the instrumental calibration of the two polarizations, instrumental polarization leakage due to asymmetry of the primary beam response and leakage from  polarized point sources \citep{Asad15,van2018,jais22}.  Faraday rotation in the magnetized plasma causes a  phase difference between the  left and right circularly polarised components \citep{Smirnov11},  and this  also can contribute to the difference in the Cross and Total $C_{\ell}(\Delta \nu)$. The DGSE  contribution  decreases   as we move to larger $\ell$. In Figure~\ref{fig:cldnu}  we see that  differences between the Total and Cross estimates of $C_{\ell}(\Delta \nu)$  go down as we move to larger $\ell$ bins.  This supports the picture where a part of the difference between the Total and Cross estimates may be attributed to the DGSE. However, this does not explain why the sharp de-correlation around $\Delta \nu=0$ and the rapid oscillations are mitigated  for the Cross estimator. This possibly has to do with gain calibrations errors and other systematics which could be uncorrelated for the two polarizations. 

\section{The Cylindrical PS}
\label{sec:cylps}

Under the flat sky approximation,  $P(k_{\perp}, k_{\parallel})$ the cylindrical power spectrum of the $21$-cm brightness temperature fluctuations $\delta T_{\rm b} (\hat{\mathbf{n}},\,\nu)$  is related to the MAPS \cl{} through a Fourier transform along the LoS \citep{KD07},
\begin{equation}
    C_{\ell}(\Delta\nu) = \frac{1}{\pi r^2} \int_{0}^{\infty}  d k_{\parallel} \cos(k_{\parallel}r^{\prime}\Delta\nu) P(k_{\perp}, k_{\parallel})
    \label{eq:cl_Pk}
\end{equation}
where $k_{\parallel}$ and $k_{\perp}=\ell/r$ are the parallel and perpendicular to the LoS components of $\mathbf{k}$ respectively. The comoving distance $r$ and  its derivative with respect to frequency $r^{\prime}=dr/d\nu$ which are evaluated at the reference frequency $\nu_{c}=432.8\,{\rm MHz}$ $(z=2.28)$, have values  $5703\,{\rm Mpc}$ and $9.85\,{\rm Mpc/MHz}$ respectively.  

We use a maximum likelihood estimator to estimate the PS $P(k_{\perp a},k_{\parallel m})$ from the measured  $C_{\ell_a }(\Delta\nu_n)$, where $n,\,m\, \in \, [0,N_{E}-1]$ and  $N_{E}$ is the number of frequency separations used in the PS estimation. In matrix notation,
\begin{equation}
    C_{\ell_a}(\Delta\nu_n)=  \sum_{m} \textbf{A}_{nm} P(k_{\perp a}, k_{\parallel m})
   + [\textrm{Noise}]_{n}
\label{eq:CL_data}
\end{equation}
where $\textbf{A}_{nm}$ are the components of the $N_{E} \times N_{E}$  Hermitian matrix $\textbf{A}$ containing the coefficients of the Fourier transform, $[\textrm{Noise}]_{n}$ is an additive noise associated with each estimated $C_{\ell_a}(\Delta\nu_n)$.   The maximum likelihood estimate of $P(k_{\perp a}, k_{\parallel m})$ is given by, 
\begin{equation}
    P(k_{\perp a}, k_{\parallel m})   =  \sum_n \left[ \left(\textbf{A} ^{\dagger} \textbf{N}^{-1} \textbf{A}\right)^{-1} \textbf{A}^{\dagger} \textbf{N}^{-1} \right]_{mn} \mathcal{W}(\Delta\nu_n) \, C_{\ell_a}(\Delta\nu_n)
    \label{eq:MLE_PS}
\end{equation}
where $\textbf{N}$ is the noise covariance matrix and `$\dagger$' denotes  the Hermitian conjugate.  Note that we have applied a Blackman-Nuttall (BN; \citealt{nut81}) window function $\mathcal{W}(\Delta\nu_n)$, normalized at $\Delta\nu = 0$, to reduce the ringing artefacts (ripples) which otherwise appears in the PS due to the discontinuity in $C_{\ell}(\Delta\nu)$ at the band edges. 

We have estimated the noise covariance matrix $\textbf{N}$ through multiple realizations ($50$ in this work)  of `noise-only' simulations of the measured complex visibilities. The random noise is assumed to follow a Gaussian distribution with zero mean and standard deviation  $ \sigma_{N}=0.43 \, \rm{Jy}$  which is estimated from the real (or imaginary) part of the measured visibility data.  Note that this assumption implies $\textbf{N}$ to be diagonal.  Further,  we have also used these simulations to estimate  $\delta P_{N}(k_{\perp}, k_{\parallel} )$  the   system noise contribution to the uncertainty in the estimated PS. An analysis of the noise statistics for $ P(k_{\perp }, k_{\parallel})$, presented later in this section, leads us to believe that the actual noise level for  the data  is approximately $4.77$ times larger than that obtained from these system noise only simulations, and we have accounted for this by scaling up all the noise  predictions by this factor.

\begin{figure}
	\includegraphics[width=\columnwidth]{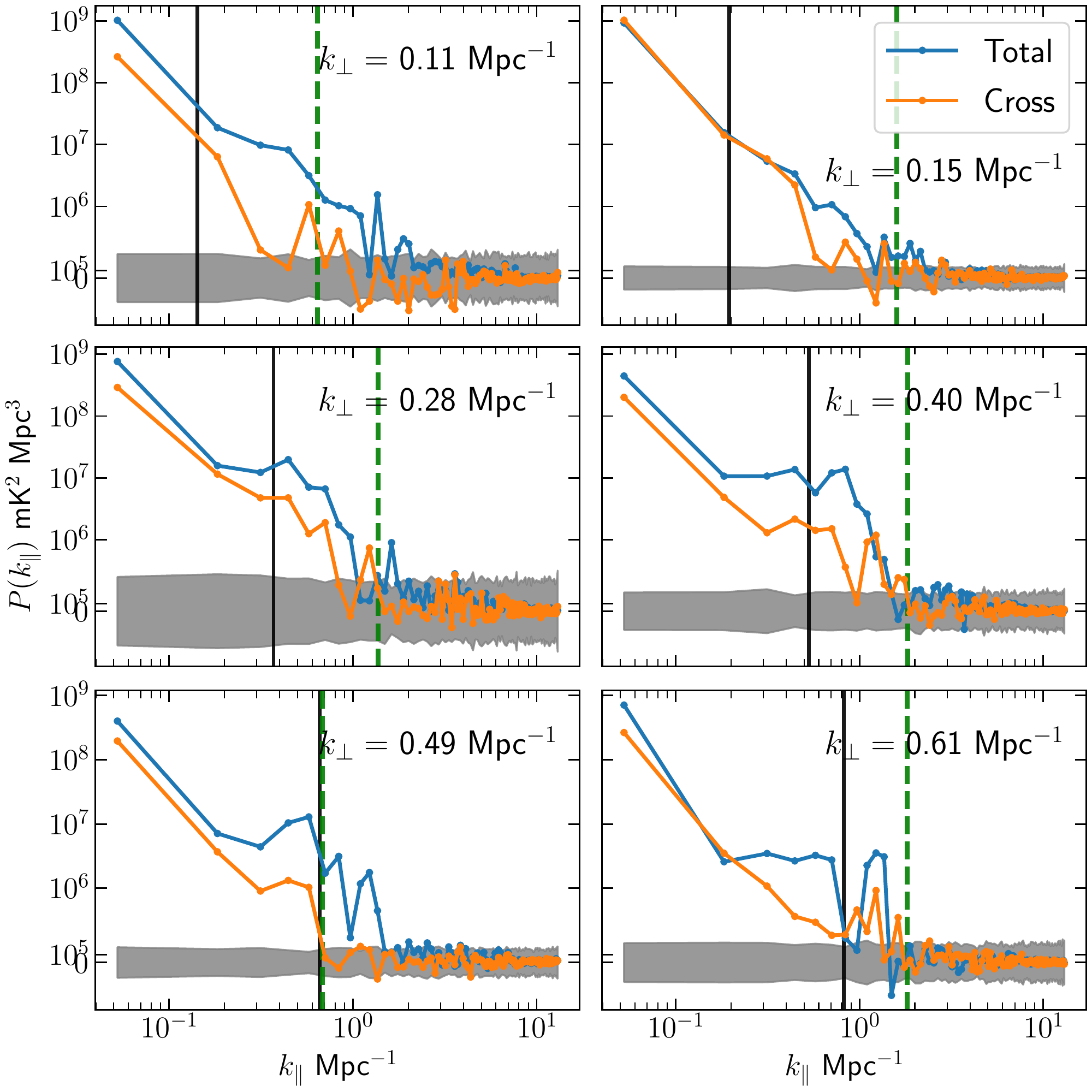}
    \caption{The Total (blue) and Cross (orange) cylindrical PS $P(k_{\perp},k_{\parallel})$ as a function of $k_{\parallel}$ for  different values of $k_{\perp}$.  The  grey shaded regions show the $3\sigma$ error bars for the Cross PS. The vertical lines show the $[k_{\parallel}]_{H}$ (black solid) and the TW ($21$-cm window) boundary (green dashed) for the respective $k_{\perp}$-bins of the Cross $P(k_{\perp},k_{\parallel})$. }
    \label{fig:pkslice}
\end{figure}

Figure~\ref{fig:pkslice} shows  both the  Total and Cross $P(k_{\perp}, k_{\parallel})$  as a function of $k_{\parallel}$, where the different panels correspond to  fixed values of $k_{\perp}$ which   are  in direct correspondence to  the  panels  in Figure~\ref{fig:cldnu}. The grey shaded regions  show the  $3 \sigma$ error bars for the Cross $P(k_{\perp}, k_\parallel{})$ and  the   black solid lines shows the theoretically predicted foreground wedge boundary  $  [k_{\parallel}]_{H} = (r/r^{\prime}\nu_{c}) k_{\perp} $ which corresponds to the foreground contribution from a  source located at the horizon. For each value of $k_{\perp}$,  we have visually inspected the Cross $P(k_{\perp}, k_\parallel{})$   and identified the region which is relatively free of foreground contamination.  We refer to this region as the `$21$-cm  window' (TW) whose boundary is demarcated  by  the  green dashed line. In the subsequent discussion we refer to the  $(k_{\perp}, k_\parallel{})$ modes  complementary to the TW (i.e., from $k_{\parallel} = 0$  to the green dashed line) as the foreground  (FG) modes. We further refer to the region  within $[k_{\parallel}]_{H}$ and the   green dashed line as the buffer.

Considering the Total PS, as noted in \citetalias{P22}, the amplitude of the PS starts with a high value $(\sim 10^9 \, {\rm mK}^2 \, {\rm Mpc}^3)$ at $k_{\parallel} = 0$ and falls with increasing $k_{\parallel}$ and nearly flattens out at $k_{\parallel} \sim 0.2-0.8 \, \rm{Mpc}^{-1}$. The amplitude then  rises slightly in a few $k_{\parallel}$ bins  just beyond  $[k_{\parallel}]_{H}$ and then again falls to $\sim 10^5  \, {\rm mK}^2 \, {\rm Mpc}^3$ at $k_{\parallel} > 1-2 \, \rm{Mpc}^{-1}$ where it oscillate between positive and negative values which are comparable with the  noise.  The Cross PS also shows a similar feature in all the $k_{\perp}$ bins but with a comparatively lower amplitude  throughout the entire $k_{\parallel}$ range. This is particularly noticeable  in the near flat region where  the Cross PS has an amplitude that is an order of magnitude smaller. Further,  it reaches the noise level  at a relatively lower  $k_{\parallel}$ $(k_{\parallel} \sim 0.8-1 \, \rm{Mpc}^{-1})$ as compared to  the total PS.

The features in $P(k_{\perp},k_{\parallel})$ (Figure~\ref{fig:pkslice}) are directly related to the features seen in the MAPS \cl{} (Figure~\ref{fig:cldnu}). A sharp variation in \cl{} yields a smooth variation in $P(k_{\perp},k_{\parallel})$, and this is why the Total PS shows a wider flat region compared to the Cross PS, and it also reaches the noise level slower than the Cross PS. This feature is markedly visible in the last four $k_{\perp}$-bins which correspond to the last four $\ell$-bins of Figure~\ref{fig:cldnu}. In these bins, the Total MAPS decorrelate faster than the Cross MAPS, and so the Total PS goes to noise level much slower than the Cross PS.

The high values of the PS beyond $[k_{\parallel}]_{H}$ is related to the rapid oscillations we see in the MAPS. The amplitude of the PS depends on the oscillation amplitude in MAPS, whereas, the oscillation period  points to the $k_{\parallel}$ mode where the power corresponding to the oscillation arises. The oscillations with larger amplitudes and small periods thus show up as the spikes in the PS at the larger $k_{\parallel}$ modes. We have seen in Figure~\ref{fig:cldnu} that the oscillation amplitude in the Cross MAPS is much smaller than in the Total MAPS and this is why the amplitude of the spikes is much smaller in the Cross PS. 

\begin{figure}
	\includegraphics[width=\columnwidth]{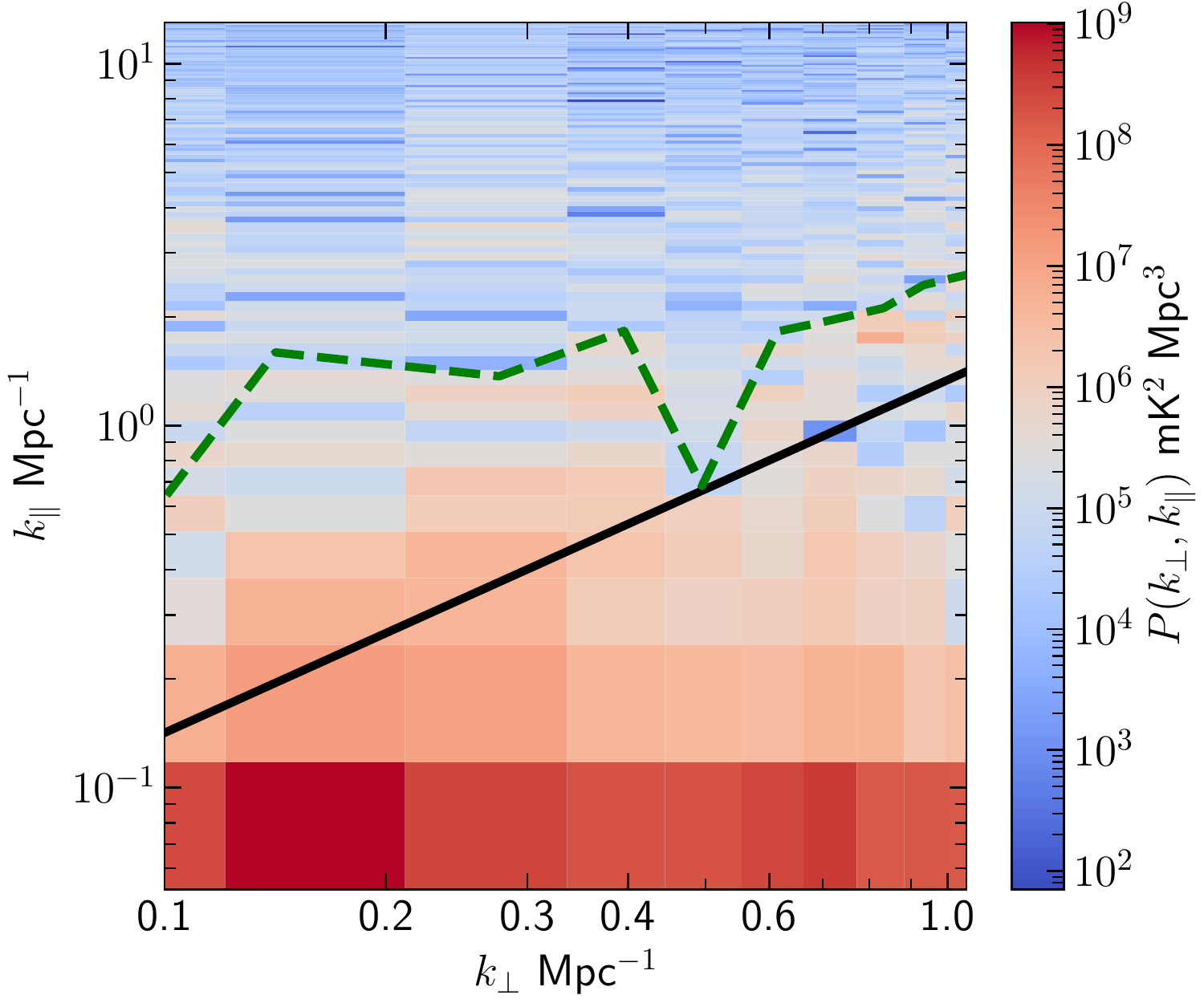}
    \caption{The Cross cylindrical power spectra $\mid P(k_{\perp},k_{\parallel}) \mid $.  Here the black solid and green dashed 
    lines denote $[k_{\parallel}]_{H}$ and the TW boundary respectively. The region above the green dashed line was identified as being relatively free of foreground contamination, and used for  spherical binning.}
    \label{fig:pscyl}
\end{figure}

We have shown the Cross PS heatmap $\mid P(k_{\perp},k_{\parallel}) \mid$ in Figure~\ref{fig:pscyl}. The  black solid line and the green dashed line denote the wedge and the buffer boundary respectively. We see that most of the power lies inside the wedge boundary where the PS vary $\sim 10^7 - 10^9 \, \rm{mK}^2\,\rm{Mpc}^3$. There is considerable foreground leakage in the buffer region where the PS vary $\sim 10^5 - 10^7 \, \rm{mK}^2\,\rm{Mpc}^3$. The buffer boundary is chosen by inspecting the $1$D slices (Figure~\ref{fig:pkslice}) in each $k_{\perp}$-bins. Considering the first $k_{\perp}$-bin we have chosen a buffer of $\sim 0.5 \, \rm{Mpc}^{-1}$. We have chosen a relatively larger buffer ($0.8 - 1.2 \, \rm{Mpc}^{-1}$) in the subsequent bins which show additional leakage barring the $5$th bin $(k_{\perp} = 0.49 \, \rm{Mpc}^{-1})$ which looks clean beyond the wedge boundary. 

It is necessary to ensure that the power in the TW is either strictly positive or consistent with noise. To ensure that our PS estimates are free from negative systematics, we study the quantity $X$, which is the ratio between the estimated cylindrical PS $P(k_{\perp},k_{\parallel})$ and the  statistical fluctuation  $\delta P_N(k_{\perp},k_{\parallel})$ expected due to the  system noise,
\begin{equation}
    X=\frac{P(k_{\perp},k_{\parallel})}{\delta P_{N}(k_{\perp},\,k_{\parallel})}.
    \label{eq:xstat}
\end{equation}
We note that $\delta P_N(k_{\perp},k_{\parallel})$  in equation~(\ref{eq:xstat}) does not include the factor of $4.77$ which was mentioned earlier. We expect  $X$ to have a symmetric  distribution with zero mean and unit standard deviation if the 
values of $P(k_{\perp},k_{\parallel})$ are entirely due to the system noise contribution. 

\begin{figure}
	\includegraphics[width=\columnwidth]{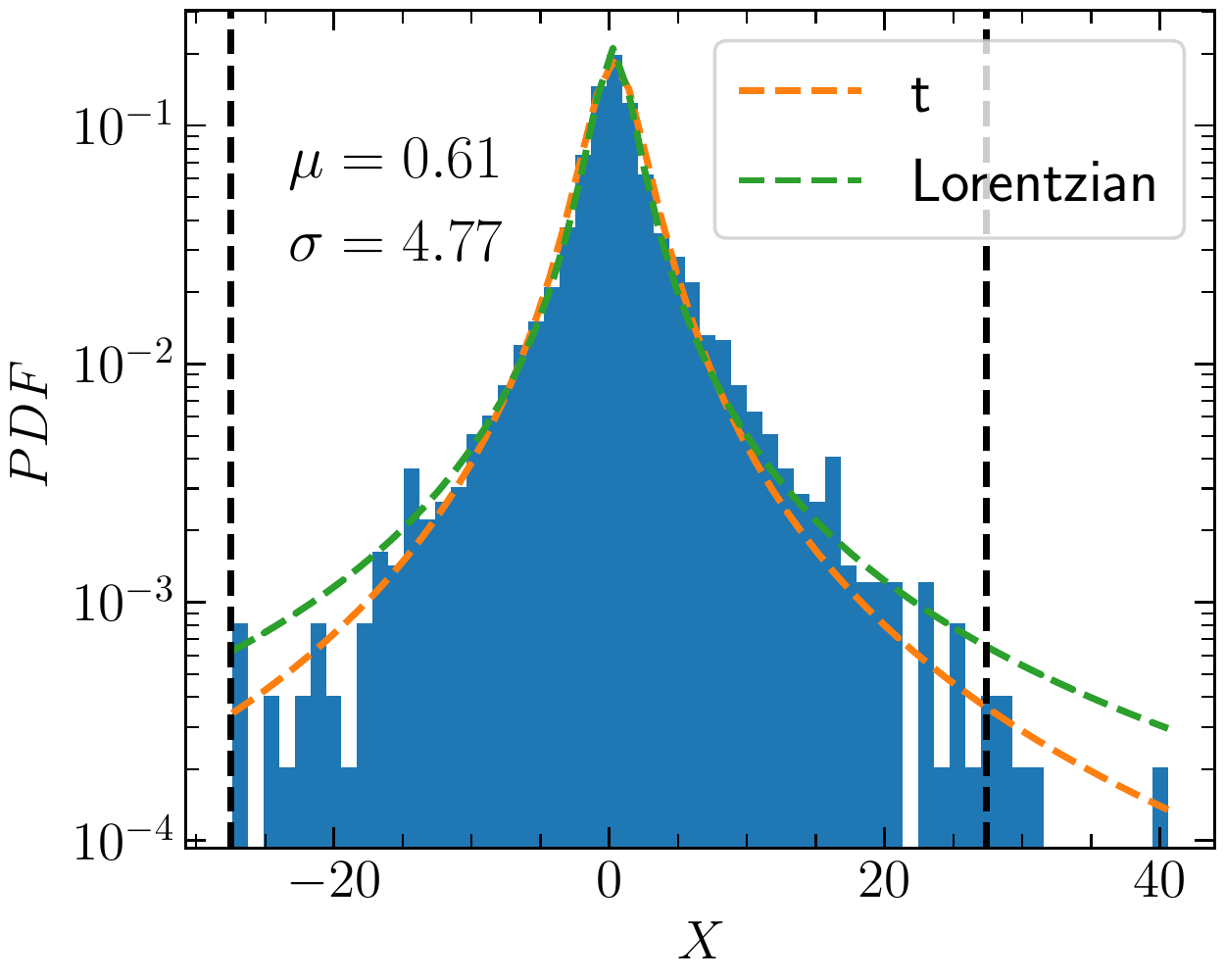}
    \caption{The histogram of the variable $X=\frac{P(k_{\perp},\,k_{\parallel})}{\delta P_{N}(k_{\perp},\,k_{\parallel})}$ is shown. The orange and green dashed curves show the fit with t and Lorentzian distributions, respectively.}
    \label{fig:noisestat}
\end{figure}

Figure~\ref{fig:noisestat} shows the histogram of $X$. We see that bulk of the data points $(99.63\%)$ in the histogram lie in the central $ \mid X \mid \le 30$ region, which we delimit by the vertical black-dashed lines. The probability density function (PDF)  is mostly symmetric in the central region with a positive mean $\mu=0.61$ and a standard deviation $\sigma_{Est}=4.77$. We do not see negative outlier values of $X$ beyond the central region. The standard deviation $\sigma_{Est} > 1$ suggests that the  statistical fluctuation in $P(k_{\perp},k_{\parallel})$ are underestimated by the system noise only simulations. This excess is possibly due to artefacts from  calibration errors, inaccurate point source subtraction and RFI. As mentioned earlier, we have scaled up all the error estimates by a factor of $4.77$ to account for this.

We have seen in \citetalias{P22} that a t-distribution adequately describes the $X$ statistics of the Total PS near the central region, but it fails to fit the positive tail. Here also, we find the t-distribution (orange dashed line) to under-fit the tail of the histogram. However, a Lorentzian distribution (green dashed line) seems to represent the statistics better. Also note that the positive tail is quite shorter for the  Cross PS as compared to the Total PS (\citetalias{P22}). This shorter positive tail suggests that  
we have a cleaner TW region with less foreground contamination for the Cross PS in comparison to the Total PS.  We have used all the $(k_{\perp},k_{\parallel})$ modes in the TW for spherical binning which we describe in Section~\ref{sec:BMLE}.

\section{The Spherical PS}
\label{sec:BMLE}
In this section we utilize  an important feature which distinguishes the redshifted $21$-cm signal from the foregrounds. This arises from  the fact that the $21$-cm signal traces out the three-dimensional distribution of a cosmological  density field.  Like all cosmological density fields, we expect the $21$-cm signal also to be statistically isotropic in three-dimensional space, i.e. its clustering properties depend only on the length of the spatial separation irrespective of the  orientation with respect to the plane of the sky and the  LoS directions. The  allows us to quantify the $21$-cm  signal using  the spherical PS $P_T(k)$, where $k = \sqrt{k_{\perp}^2 + k_{\parallel}^2}$. We note that this isotropy is broken by redshift space distortion \citep{BA5}. While it is also possible to include this effect in our analysis, we have chosen to ignore it for the present work. The $21$-cm MAPS $[C_{\ell}(\Delta \nu)]_T$, which is  related to $P_T(k)$ through equation~(\ref{eq:cl_Pk}), is expected to encode this isotropy through its $\ell$ and $\Delta \nu$ dependence.  This distinguishes  $[C_{\ell}(\Delta \nu)]_T$ from the other sources for which frequency separation $\Delta \nu$ does not correspond to a spatial separation. 

\begin{figure}
    \centering
	\includegraphics[width=\columnwidth]{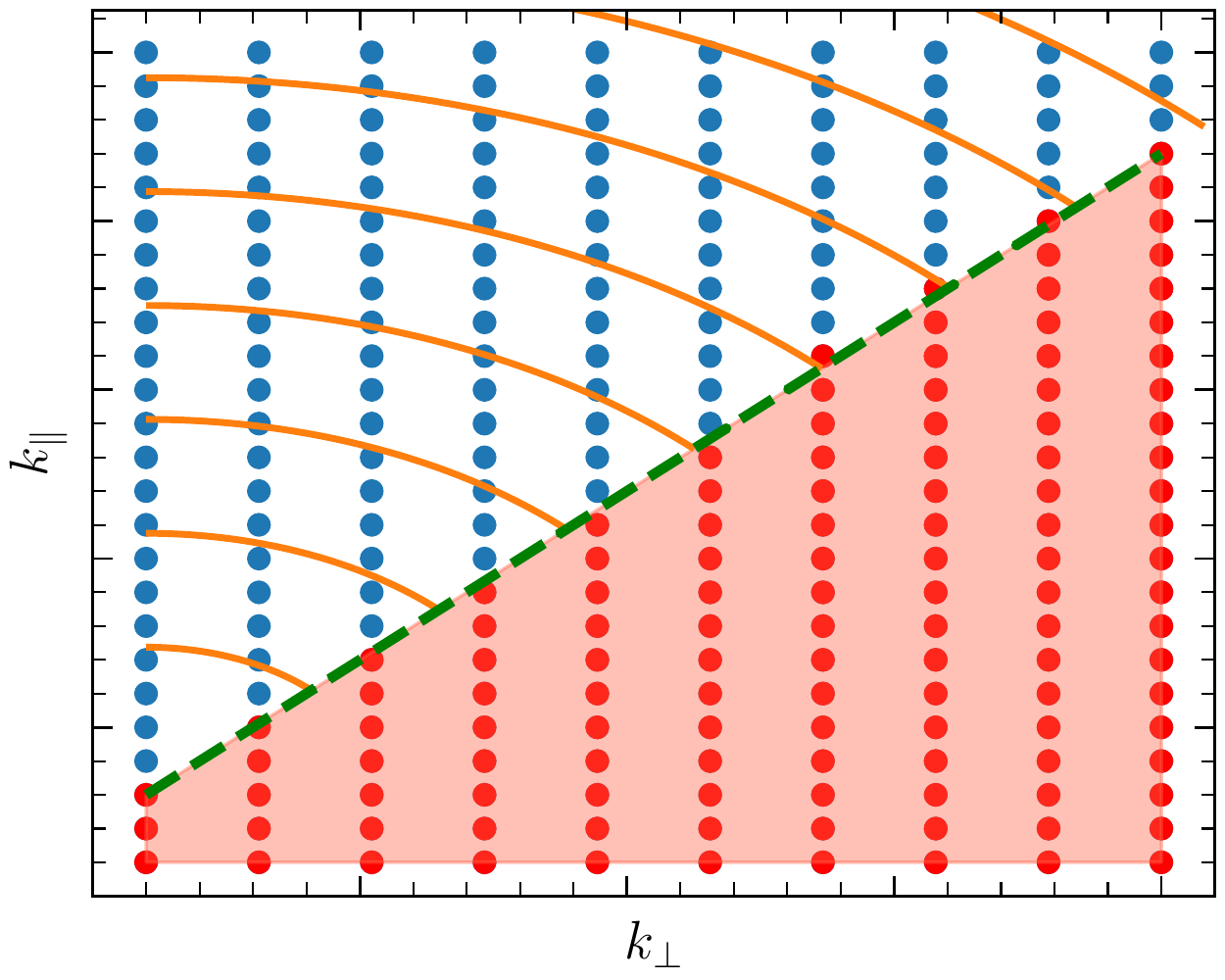}
    \caption{A schematic diagram to explain the spherically binning MLE. Each  filled  circle in the diagram represents a $(k_{\perp}, k_{\parallel})$ mode. The FG modes are shown in red, whereas the modes in the TW are shown in blue. The green dashed curve demarcates the boundary of the TW which is divided into equally spaced logarithmic bins shown by the orange spherical arcs.}
    \label{fig:diagram_mle}
\end{figure}

Here we have modelled the measured $C_{\ell}(\Delta \nu)$ as 
\begin{equation}
    C_{\ell_a}(\Delta\nu_n) = \left[C_{\ell_a}(\Delta\nu_n)\right]_{FG} + \left[C_{\ell_a}(\Delta\nu_n)\right]_{T} + \left[C_{\ell_a}(\Delta\nu_n)\right]_{R}
	\label{eq:Cmodel}
\end{equation}
and used this to estimate the spatially isotropic component $[C_{\ell_a}(\Delta\nu_n)]_{T}$.  As mentioned earlier, $ C_{\ell_a}(\Delta\nu_n) $  is dominated by  $[C_{\ell_a}(\Delta\nu_n)]_{FG}$   the foreground contribution. The foregrounds are expected to have a smooth frequency dependence, and the $\Delta \nu$ dependence arises mainly due to instrumental effects like baseline migration (\citealt{Morales_2012,hazelton2013}).   Considering Figure~\ref{fig:pscyl},  we have identified  a region of $(k_{\perp},k_{\parallel})$  plane where the modes are foreground dominated (FG modes). Further, the complementary region, referred to as the $21$-cm window (TW), was identified as being relatively free of foreground contamination.    Here we have assumed that $[C_{\ell_a}(\Delta\nu_n)]_{FG}$ can be entirely quantified in terms of the FG  modes as 
\begin{equation}
    \left[C_{\ell_a}(\Delta\nu_n)\right]_{FG} = \sum_{m} A_{nm}  \left[ P(k_{\perp a}, k_{\parallel m}) \right]_{FG}\,  
	\label{eq:CFG}
\end{equation} 
and we have excluded these modes for estimating the $21$-cm signal. The modes within this region are schematically represented by the red points in Figure~\ref{fig:diagram_mle}, where  the green dashed line denotes the boundary of the TW. 

We have  used only the  TW modes  to estimate the $21$-cm signal.  
Here  we have assumed  spatial  isotropy and divided the TW into spherical bins (labelled $i=1, ...,{\rm NBin}$) which are  shown schematically in Figure~\ref{fig:diagram_mle}. We use  $[P(k_i)]_T$  to denote the value of the spherical PS corresponding to the $i$-th bin. We have modelled the $21$-cm signal as 
\begin{equation}
    \left[C_{\ell_a}(\Delta\nu_n)\right]_{T} =   \sum_{i} B_i(a , n) \left[ P(k_i) \right]_{T} 
	\label{eq:CTPK}
\end{equation}
with $B_i(a , n) = \sum_{m} A_{nm}$ where this sum is over the  $(k_{\perp a},k_{\parallel m})$ modes which are within the $i$-th bin. Note that we have dropped the subscript `T' in $[P(k_i)]_T$, and simply denote it as $P(k)$ when there is no ambiguity.

Considering equation~(\ref{eq:Cmodel}),  $ [C_{\ell}(\Delta\nu)]_{R}$ refers to the residual MAPS i.e. the component  of $C_{\ell}(\Delta \nu)$ which is not included in  the foregrounds or the isotropic $21$-cm signal. Noise, systematics and foreground leakage are possible factors which contribute to $ [C_{\ell}(\Delta\nu)]_{R}$ \citep{jais2020}. Ideally, we expect $ [C_{\ell}(\Delta\nu)]_{R}$ to be consistent with our noise estimates,  and we define  chi-square $(\chi^2)$ as
\begin{equation}
     \chi^2 = \sum_{a,n,m} \left[C_{\ell_a}(\Delta\nu_n)\right]_R \, \textbf{N}^{-1}_{nm} \,  \left[C_{\ell_a}(\Delta\nu_m)\right]_R
	\label{eq:chi2}
\end{equation}
where $\textbf{N}$ is the noise covariance matrix introduced  in equation~(\ref{eq:MLE_PS}).
Our model for the measured $C_{\ell}(\Delta \nu)$ now has $ [ P(k_{\perp a}, k_{\parallel m})]_{FG}$ and $[P(k_i)]_T$ as parameters. We have maximized the likelihood $\mathcal{L} \propto \exp{(-\chi^2 /2)}$ with respect to the parameters in order to determine the best fit parameter values. We have also used this likelihood analysis to obtain  error estimates for the best fit parameter values. 

\begin{figure}
	\includegraphics[width=\columnwidth]{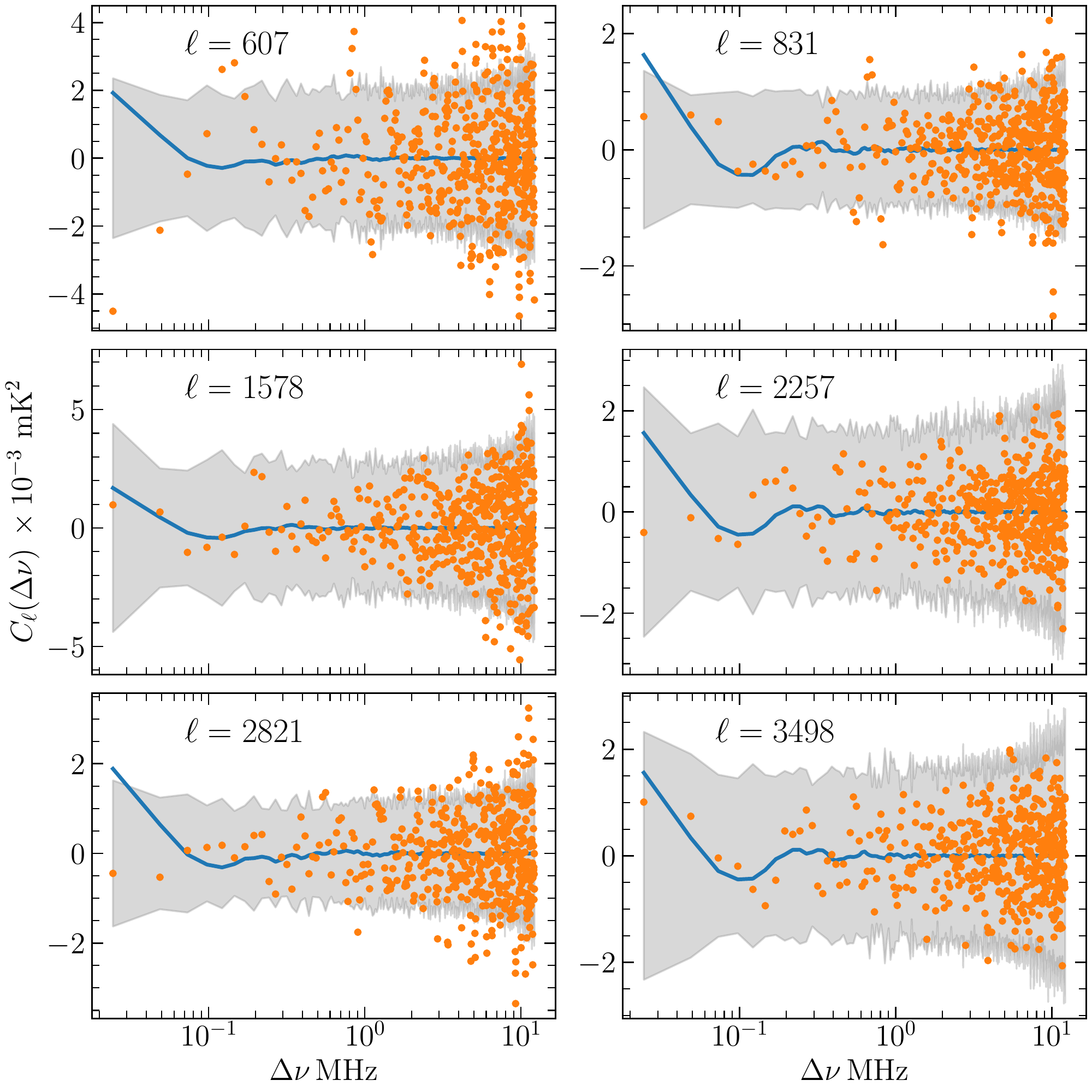}
    \caption{For the first $6$ $\ell$ bins, this shows $[C_{\ell}(\Delta\nu)]_{T}$ (solid blue lines)  and $[C_{\ell}(\Delta\nu)]_{R}$ (orange dots) corresponding to the best fit $ \left[ P(k_i) \right]_{T}$ .     The grey shaded regions show $2\,\sigma$ error bars  for  the measured \cl{}. }
    \label{fig:mle_PK}
\end{figure}

Here we have used a total $5000$ measured data points $C_{\ell_a}(\Delta\nu_n)$ corresponding to $10$ $\ell$-bins and $500$ frequency separations $\Delta\nu$  to 
obtain maximum likelihood estimates  for  a total $664$ parameters,  of which $656$ are the FG modes $[ P(k_{\perp a}, k_{\parallel m}) ]_{FG}$  and the remaining  $8$  are the $[ P(k_i) ]_{T}$ corresponding to the $8$ spherical $k$ bins which span  $0.804 < k < 11.892 \, \rm{Mpc}^{-1}$. We find that the goodness-of-fit parameter (reduced-$\chi^2$) has a value  $1.21$ which indicates that our model provides an adequate fit for the measured $C_{\ell_a}(\Delta\nu_n)$ and  the residual $[C_{\ell_a}(\Delta\nu_n)]_R$ is roughly consistent with noise. We have used the best fit  $[ P(k_{\perp a}, k_{\parallel m}) ]_{FG}$ and $[ P(k_i) ]_{T}$ in  equations~(\ref{eq:CFG}) and (\ref{eq:CTPK}) to recover $[C_{\ell_a}(\Delta\nu_n)]_{FG}$ and  $[C_{\ell_a}(\Delta\nu_n)]_{T}$ respectively. The measured \cl{} is foreground dominated, and we find that the recovered $[C_{\ell_a}(\Delta\nu_n)]_{FG}$ closely matched the measured \cl{} shown earlier in Figure~\ref{fig:cldnu}. 
The recovered $[C_{\ell_a}(\Delta\nu_n)]_{T}$  are shown in the different panels of Figure~\ref{fig:mle_PK}. The residuals $[ C_{\ell}(\Delta\nu) ]_{R}$ and the $2\sigma$ errors due to noise are shown using the  orange dots and the grey shaded regions respectively. We see that, in all the $\ell$-bins shown here, $[C_{\ell_a}(\Delta\nu_n)]_{T}$ varies  within $0-0.002 \, \rm{mK}^2$ and the values  lie within the predicted $2\sigma$ noise levels. Further, in all cases the recovered $[C_{\ell_a}(\Delta\nu_n)]_{T}$  has maximum value at $\Delta \nu=0$,  the value decreases with increasing $\Delta \nu$ and is close to $0$ for large $\Delta \nu$.  We also notice some oscillatory features  in  $[C_{\ell_a}(\Delta\nu_n)]_{T}$ which reflect the fact that some of the $(k_{\perp a}, k_{\parallel m})$ modes (in the FG region) were excluded when calculating  $[C_{\ell_a}(\Delta\nu_n)]_{T}$. Although there are some outliers at large $\Delta \nu$ (possibly due to larger cosmic variance), the residual $[C_{\ell}(\Delta\nu)]_{R}$ are found to be  largely  consistent with the $0\pm2\sigma$ noise levels. 

\begin{figure}
	\includegraphics[width=\columnwidth]{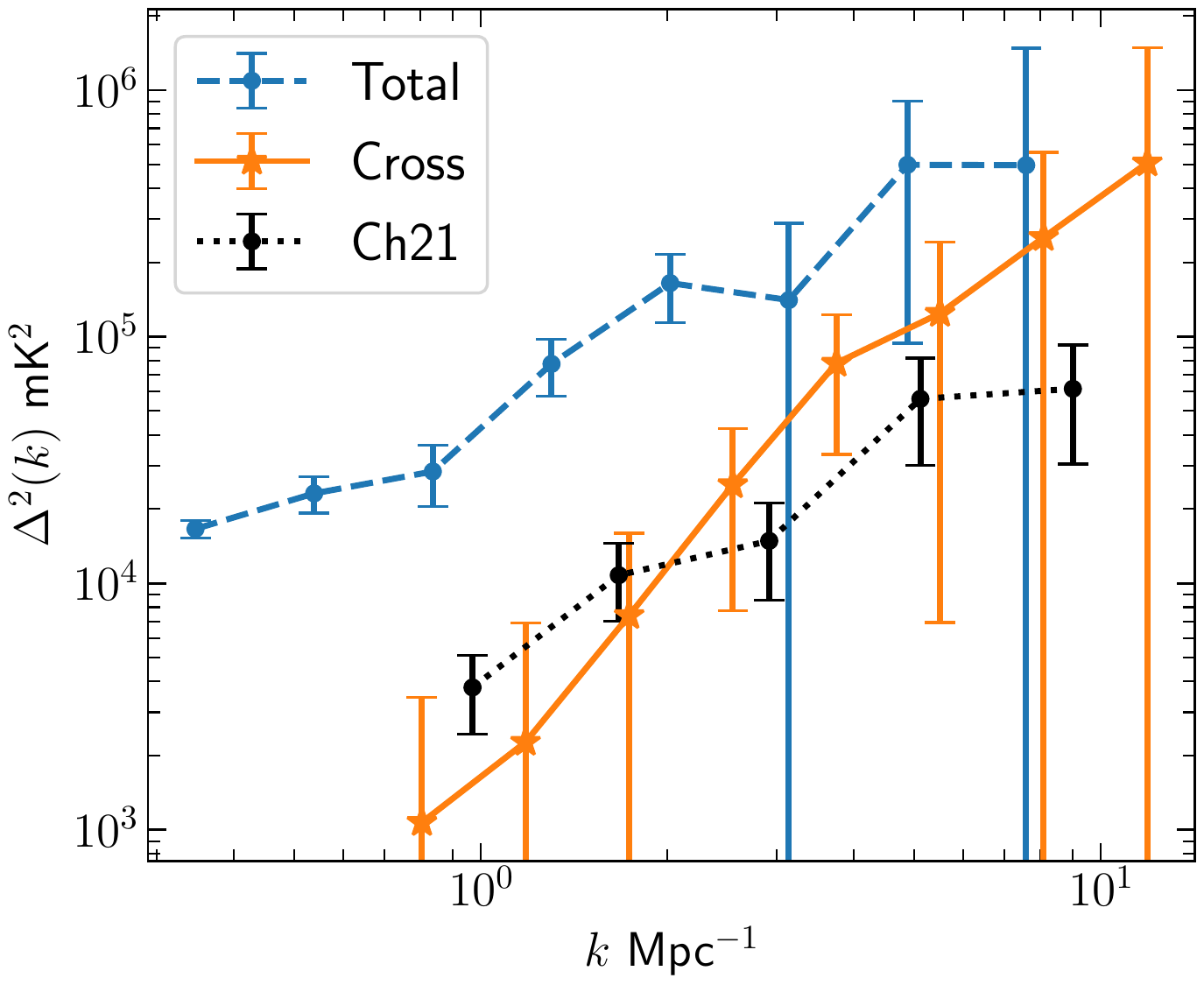}
    \caption{The mean squared brightness temperature fluctuations $\Delta^2(k)$ along with $2 \sigma$ error bars. The orange asterisks show the results from the present work (Cross), while the blue dashed and  black dotted lines show the results from   \citetalias{P22} (Total)  and \citetalias{Ch21},  respectively.}
    \label{fig:pssph}
\end{figure}

We have used the  best fit  $[P(k)]_T$ values to calculate the mean squared brightness temperature $\Delta^2(k)\equiv {k^{3}}P(k)/{2\pi^{2}}$  shown by the orange asterisks in Figure~\ref{fig:pssph} along with the corresponding $2\sigma$ error bars. The $\Delta^2(k)$ values, $\sigma$  and the signal-to-noise ratio (SNR, $\Delta^2(k)/\sigma$)  are tabulated for different $k$-bins in Table~\ref{tab:ul_MLE}. We find that  $\Delta^2(k) >0$ for all the $k$ bins.  The  $\Delta^2(k)$ values in the first $3$ and the last $2$ $k$ bins are consistent with noise at the $0\pm 2\sigma$ level,  whereas  it  is slightly  above $2\sigma$  in the sixth  bin. We interpret the $\Delta^2(k)$ estimated in these $6$ bins as  arising from noise. The values of $\Delta^2(k)$ in the fourth and fifth bins  exceed $0+2\sigma$,  but are within  $0+3\sigma$ and $0+5\sigma$ respectively. The  $\Delta^2(k)$ estimated in these two bins may have a contribution from residual systematics or foreground leakage.

\begin{table}
    \centering
    \caption{The mean squared brightness temperature fluctuations $\Delta^2(k)$ and the corresponding statistical error predictions $\sigma$ for different $k$-bins. The $2\,\sigma$ upper limits  $\Delta_{UL}^{2}(k)=\Delta^{2}(k)+2\,\sigma$ and corresponding $[\Omega_{\HI}b_{\HI}]_{UL}$ values are also provided.}
        \begin{tabular}{cccccc}
        \hline
        \hline
        $k$ & $\Delta^2(k)$ & $1\sigma$ & SNR & $\Delta_{UL}^{2}(k)$ & $[\Omega_{\HI}b_{\HI}]_{UL}$ \\
        Mpc$^{-1}$ & (mK)$^2$ & (mK)$^2$ & & (mK)$^2$  &\\
        \hline
        $0.804$ &  $(32.75)^2$ & $(34.42)^2$ & $0.905$ & $(58.67)^2$ & $0.072$\\
        $1.181$ &  $(47.64)^2$ & $(48.19)^2$ & $0.977$ & $(83.15)^2$ & $0.089$\\
        $1.736$ &  $(86.05)^2$ & $(65.31)^2$ & $1.736$ & $(126.24)^2$ & $0.121$\\
        $2.551$ &  $(158.47)^2$ & $(93.18)^2$ & $2.892$ & $(206.11)^2$ & $0.177$\\
        $3.748$ &  $(279.47)^2$ & $(149.74)^2$ & $3.483$ & $(350.64)^2$ & $0.273$\\
        $5.507$ &  $(352.76)^2$ & $(242.38)^2$ & $2.118$ & $(491.87)^2$ & $0.350$\\
        $8.093$ &  $(502.73)^2$ & $(391.31)^2$ & $1.651$ & $(747.65)^2$ & $0.490$\\
        $11.892$ &  $(712.14)^2$ & $(698.77)^2$ & $1.039$ & $(1218.07)^2$ & $0.589$\\
        \hline
    \end{tabular}
    \label{tab:ul_MLE}
\end{table}

Considering the $\Delta^2(k)$ values, we find that $\Delta^2(k)$ has the smallest value  $(32.75)^2 \, \rm{mK}^2$ at the lowest $k$-bin where $k=0.804 \, \rm{Mpc}^{-1}$. The values of $\Delta^2(k)$ as well as the $1\sigma$ errors are found to increase with increasing $k$ as a power-law $k^{n}$, where the exponent $n\sim2.5$ for $\Delta^2(k)$ and $n\sim2.4$ for $\sigma(k)$ respectively. We have used the estimated $\Delta^2(k)$ and the $\sigma$ values to place $2 \sigma$ upper limits $\Delta_{UL}^{2}(k) = \Delta^{2}(k) + 2\sigma$ on the $21$-cm brightness temperature fluctuations at different $k$ values. The  $2\sigma$  upper limits $\Delta_{UL}^2(k)$ are also tabulated in Table~\ref{tab:ul_MLE}. We find the tightest constraint on the upper limit to be $\Delta_{UL}^2(k)\leq (58.67)^2 \, \rm{mK}^2$ at $k = 0.804 \, \rm{Mpc}^{-1}$.

Figure~\ref{fig:pssph} also shows (blue dashed line) the results from \citetalias{P22} where we have used the Total TGE as against the Cross TGE used here. We find that the present analysis shows significant improvement over \citetalias{P22} throughout the entire $k$-range. Particularly  near $ k \sim 1 \,\rm{Mpc}^{-1}$, the value of $\Delta^{2}(k)$ is nearly $20$ times smaller in the present analysis as compared to \citetalias{P22} whereas this factor is around $4-6$ for the larger $k$ bins. Comparing the upper limits, we had 
$\Delta_{UL}^2(k)\leq (133.97)^2 \, \rm{mK}^2$ at $k = 0.347 \, \rm{Mpc}^{-1}$ in \citetalias{P22} which is tightened to  $\Delta_{UL}^2(k)\leq (58.67)^2 \, \rm{mK}^2$ at $k = 0.804 \, \rm{Mpc}^{-1}$ in the present work. Note that the lowest $k$ bin here is somewhat larger than that in  \citetalias{P22}. We have also compared our findings with \citetalias{Ch21} who have conducted a multi-redshift analysis of the same observational  data after splitting it into four sub-bands, each of $8$ MHz bandwidth. The black dashed line in Figure~\ref{fig:pssph} shows the $\Delta^{2}(k)$ values from their $z=2.19$ sub-band which is the close to our analysis $(z=2.28)$.  We find that the present upper limits are close to the findings of \citetalias{Ch21} who reported $\Delta_{UL}^2(k)\leq (61.49)^2 \, \rm{mK}^2$ at $k = 1 \, \rm{Mpc}^{-1}$ at the redshift $z=2.19$. We note that the bandwidth of the data analysed here is larger than that used in  \citetalias{Ch21}.  For a nearly one-to-one comparison with  \citetalias{Ch21}, we have repeated the analysis using the same $8$ MHz bandwidth for which the results are presented in Appendix~\ref{sec:comparison}. We find that the results are very similar to those presented here.

The upper limits on the $21$-cm brightness temperature fluctuations allow us to constrain the cosmological \HI{} abundance parameter $[\Omega_{\HI}b_{\HI}]$. Here $\Omega_{\HI}$ is the comoving neutral hydrogen mass density in units of the current critical density \citep{Lanzetta95}, and $b_{\HI{}}$ is the \HI{} bias parameter. The assumption here is that the \HI{} distribution traces the  underlying matter distribution through $b_{\HI}$. This assumption allows us to express  $P_{T}(\mathbf{k})$ in terms of  $P^s_m(\mathbf{k})$  the underlying matter power spectrum in redshift space. Here we use equation (13) and (14) of \citetalias{P22} (which has been taken from \citealt{BA5}),
\begin{equation}
    P_{T}(\mathbf{k}) = \left[\Omega_{\HI} b_{\HI}\right]^{2} \bar{T}^{2} P^s_{m}(\mathbf{k})
    \label{eq:pT}
\end{equation}
with the mean brightness temperature $\bar{T}$
\begin{equation}
    \bar{T}(z) = 133 \,{\rm mK}\,(1+z)^{2} \left(\frac{h}{0.7}\right) \left(\frac{H_{0}}{H(z)}\right)
    \label{eq:tbar}
\end{equation}
and  $P^s_{m}(\mathbf{k})$ is the underlying dark matter power spectrum in redshift space for which   we have used a fitting formula \citep{Eisenstein_1998},  ignoring the effect of redshift space distortion. 

We have used the estimated  $\Delta_{UL}^{2}(k)$  to place the corresponding $2 \sigma$ upper limits   $[\Omega_{\HI}b_{\HI}]_{UL}$ which are also tabulated in Table~\ref{tab:ul_MLE}. We obtain the tightest constraint of $[\Omega_{\HI}b_{\HI}]_{UL} \le  0.072$ from the smallest bin $k = 0.804 \, \textrm{Mpc}^{-1}$. This is a factor of $3$ improvement over  \citetalias{P22} where  we were able to constrain $[\Omega_{\HI}b_{\HI}]_{UL} \le  0.23$  at $k = 0.347 \, \textrm{Mpc}^{-1}$.  \citetalias{Ch21} reported $[\Omega_{\HI}b_{\HI}]_{UL} \le  0.11$  at $k = 1 \, \textrm{Mpc}^{-1}$ which is close to the upper limit that we obtain here. 

This maximum likelihood approach of estimating $P(k)$ is different from the usual spherical binning approach (e.g. \citetalias{P22}). The maximum likelihood estimator (MLE) is robust in the presence of small numbers of somewhat larger outliers \citep{Huber}, and is optimal as we use inverse noise covariance weightage in the likelihood. We have validated the MLE in Appendix~\ref{sec:validation}. We have also carried out a consistency check on the best fit  solutions and the error estimates of MLE by sampling the posterior probability distributions of the parameters using a Markov Chain Monte Carlo (MCMC) algorithm. The details of the MCMC analysis are presented in Appendix~\ref{sec:mcmc}.

\section{Constraining \texorpdfstring{\obh{}}{Omb}}
\label{sec:omb}

In this section we consider the possibility of utilizing the entire set  of measured  $[C_{\ell_a}(\Delta\nu_n)]$ values  to directly constrain a single parameter $[\Omega_{\HI}b_{\HI}]$, without involving an intermediate step of estimating the spherical PS  $[ P(k) ]_{T}$.  Here also we have modelled the measured $ C_{\ell_a}(\Delta\nu_n)$ using equation~(\ref{eq:Cmodel}),  and used equation~(\ref{eq:CFG}) to model the foreground component $\left[C_{\ell_a}(\Delta\nu_n)\right]_{FG}$. Considering the $21$-cm signal, we have used only the TW modes  to model     $\left[C_{\ell_a}(\Delta\nu_n)\right]_{T}$  using
\begin{equation}
    \left[C_{\ell_a}(\Delta\nu_n)\right]_{T} =  \left[\Omega_{\HI} b_{\HI}\right]^{2} \bar{T}^{2}  \sum_{q} A_{nq} \, P_m(k_{\perp a}, k_{\parallel q})  
	\label{eq:CTPK_Omb}
\end{equation}
where $P_{m}(\mathbf{k})$  is the dark matter power spectrum (equation~\ref{eq:pT}) ignoring the effect of redshift space distortion. The entire $21$-cm signal is now quantified by  a single parameter  $[\Omega_{\HI}b_{\HI}]^2$. Here we have used maximum likelihood to simultaneously estimate the best fit values of  the amplitude of FG modes $[ P(k_{\perp a}, k_{\parallel m}) ]_{FG}$ and  $[\Omega_{\HI}b_{\HI}]^2$.

\begin{figure}
	\includegraphics[width=\columnwidth]{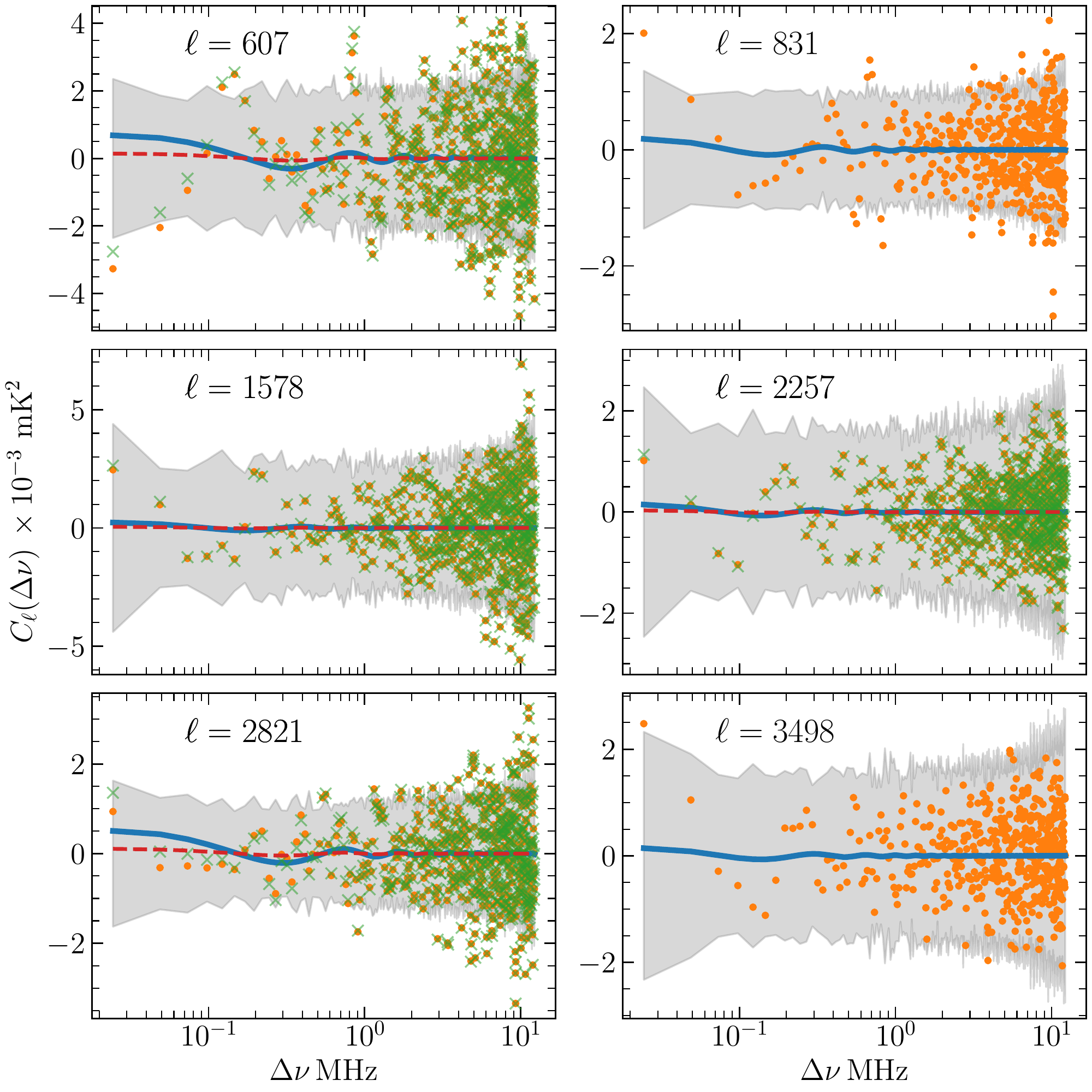}
    \caption{For the first $6$ $\ell$ bins, the figure shows the recovered $[C_{\ell}(\Delta\nu)]_{T}$ (solid blue lines), and $[C_{\ell}(\Delta\nu)]_{R}$ (orange dots) corresponding to the best fit value of $[\Omega_{\HI}b_{\HI}]^2$ obtained from Set I (Table~\ref{tab:ul_omb}). The recovered $[C_{\ell}(\Delta\nu)]_{T}$ (red dashed lines) and $[C_{\ell}(\Delta\nu)]_{R}$ (green crosses) corresponding to the best fit value of $[\Omega_{\HI}b_{\HI}]^2$ obtained from Set II are also shown. The $2\,\sigma$ error in the measured \cl{} is shown by the grey shaded regions.}
    \label{fig:mle_CL}
\end{figure}

We have considered the measured $[C_{\ell_a}(\Delta\nu_n)]$ from all the available $\ell$-bins to constrain $[\Omega_{\HI}b_{\HI}]^2$. The total number of parameters now become $657$, $656$ of which represent $[ P(k_{\perp a}, k_{\parallel m}) ]_{FG}$ and $1$ parameter for $[\Omega_{\HI}b_{\HI}]^2$.  The best fit value of $[\Omega_{\HI}b_{\HI}]^2$  is found to be $3.57 \times 10^{-3}$ with an associated $1\sigma$ uncertainty of $1.41 \times 10^{-3}$.  The  reduced $\chi^2 $ is found to to be $ 1.21$, which is the same as found in Section~\ref{sec:BMLE} where we estimated the spherical PS. The blue solid lines in the different panels of  Figure~\ref{fig:mle_CL} show the recovered $[C_{\ell}(\Delta\nu)]_T$ corresponding to the best fit value of $[\Omega_{\HI} b_{\HI}]^2$. The residuals $[ C_{\ell}(\Delta\nu) ]_{R}$ and the $2\sigma$ errors due to noise are shown using the  orange dots and the grey shaded regions respectively. We find that  the recovered $[C_{\ell}(\Delta\nu)]_{T}$ as well as  the residuals $[ C_{\ell}(\Delta\nu)]_{R}$ are roughly consistent with the $2\sigma$ noise level. The $2\sigma$ upper limit on $[\Omega_{\HI}b_{\HI}]$ is found to be $[\Omega_{\HI}b_{\HI}]_{UL} \leq 0.080$. These results, hereafter referred to as `Set I',  are tabulated in Table~\ref{tab:ul_omb}.

\begin{table}
    \centering
    \caption{$[\Omega_{\HI}b_{\HI}]^2$, its associated error due to noise, SNR, and the $2\,\sigma$ upper limits $[\Omega_{\HI}b_{\HI}]_{UL}$ are shown.}
        \begin{tabular}{cccccc}
        \hline
        \hline
        $\multirow{2}{*}{$\rm{Set}$}$ & $\multirow{2}{*}{$\ell$}$ & $[\Omega_{\HI}b_{\HI}]^2$ & $\rm{error}$ & $\multirow{2}{*}{$\rm{SNR}$}$  &  $\multirow{2}{*}{$[\Omega_{\HI}b_{\HI}]_{UL}$}$    \\
        & & $\times 10^{-4}$ & $\times 10^{-3}$ &  &  \\
        \hline
        $\rm{I}$ & $\rm{all}$ & $35.74$ & $1.41$ & $2.53$ &  $0.080$\\        
        \hline
        $\multirow{2}{*}{$\rm{II}$}$ & $617, 1578,$ & $\multirow{ 2}{*}{7.51}$ & $\multirow{ 2}{*}{1.47}$ & $\multirow{ 2}{*}{0.51}$ &  $\multirow{ 2}{*}{0.061}$\\
        & $2257 \,\rm{\&}\, 2821$ &  &  &  &  \\
        \hline
        
    \end{tabular}
    \label{tab:ul_omb}
\end{table}

We next consider the possibility of improving the constraints on $[\Omega_{\HI}b_{\HI}]^2$ by using a subset of the measured $[C_{\ell_a}(\Delta\nu_n)]$. Here we have repeated the analysis considering  various combinations of $\ell$ bins to  find that  a particular set $(\ell =617,\,1578,\,2257\,\rm{and}\,2821)$ provides better constraints. These results, hereafter referred to as  `Set II', are also tabulated in Table~\ref{tab:ul_omb}. We have modelled the $2000$ available $[C_{\ell_a}(\Delta\nu_n)]$ values using  $175$ parameters for $[ P(k_{\perp a}, k_{\parallel m}) ]_{FG}$ and $1$ parameter for $[\Omega_{\HI}b_{\HI}]^2$,  and  find the reduced $\chi^2$ to be $1.58$ which indicates an acceptable fit. The best fit value of $[\Omega_{\HI}b_{\HI}]^2$  is found to be $7.51\times 10^{-4} \pm 1.47\times 10^{-3}$ which can be attributed to noise. The dashed red lines in Figure~\ref{fig:mle_CL} shows the recovered $[C_{\ell}(\Delta\nu)]_T$ whose values are found to be close to zero throughout the $\Delta \nu$ range. The residuals $[ C_{\ell}(\Delta\nu) ]_{R}$ (green crosses) are found to be quite similar to those found for Set I. We conclude that we are able to separate out the foregrounds from the measured \cl{} in Set II. We  place  a   $2\sigma$ upper limit of $[\Omega_{\HI}b_{\HI}]_{UL} \leq 0.061$ which is better  than the upper limit obtained from Set I or from the spherical PS (Table~\ref{tab:ul_MLE}). 

Although we do not report a detection of the $21$-cm signal, we compare the upper limit with the available observational and theoretical constraints on the parameters $\Omega_{\HI}$ and $b_{\HI}$  at various redshifts in the post-reionization universe (\citealt{hamsa2015} and references therein). The \HI{} spectral stacking analysis (e.g. \citealt{rhee2016}) at  $z<0.5$, \HI{} IM experiments in cross-correlation with galaxy surveys (e.g. \citealt{chang10, masui2013, chime22}) at $z<1.3$,  observations of DLAs and sub-DLAs from quasar spectra (e.g. \citealt{prochaska04, pero2005, kanekar09, prohaska2009, Not, Zafar}) at $2<z<5.5$ estimate $\Omega_{\HI}\sim 10^{-3}$. On the other hand, various simulations (e.g. \citealt{marin2010, bagla2010, guha2012, Deb16}) indicate that $1 \le b_{\HI} \le 2$  for the redshift we have considered here. These values of  $\Omega_{\HI}$ and $b_{\HI}$ imply that our present upper limit $[\Omega_{\HI}b_{\HI}]_{UL} \leq 0.061$ is  roughly $30-60$ times larger than currently estimated values.

\section{Summary and Conclusions}
\label{sec:conclusion}

The $21$-cm intensity mapping (IM) has long been recognized as a powerful technique for efficiently mapping the large-scale structures in the universe out to high redshifts. Aiming a $21$-cm IM at $z=2.28$ in \citetalias{P22}, we considered a $24.4 \, \rm{MHz}$ bandwidth data which was taken from a $4$ nights observation of the ELAIS-N1 field at the Band 3 $(250-500 \, \rm{MHz})$ of uGMRT. We analysed the flagged, calibrated, and point source subtracted  visibility data (details in \citealt{Cha2}) with the TGE which allows us to taper the sky response to suppress the contribution from sources in the periphery of the telescope's field of view. In this work, we introduce a Cross TGE which grids the cross-polarizations (i.e., two mutually perpendicular polarizations) RR and LL of visibilities independently, and then correlates them to obtain \cl{}. We expect this to mitigate several effects like noise bias, calibration errors, etc., which affect the Total \cl{} (used in \citetalias{P22}) which combines the two polarizations.

We have compared the \cl{} estimated from the Total and the Cross estimators in Section~\ref{sec:maps}. We find that  \cl{} from both the estimators have comparable values once the DC component is subtracted out (Figure~\ref{fig:cldnu}). Considering the $\Delta\nu$ dependence, we find that the Total \cl{} sharply decorrelates within $\Delta\nu <1 \, \rm{MHz}$ and  exhibits rapid oscillations at larger $\Delta\nu$. In comparison, the Cross \cl{} decorrelates smoothly, and also with a considerably smaller oscillation amplitude. A combination of these two effects is reflected in the cylindrical PS $P(k_{\perp},k_{\parallel})$ which is evaluated from \cl{} using equation~(\ref{eq:MLE_PS}). Compared to the Total PS, we find that the Cross PS reaches the expected noise level at comparatively smaller $k_{\parallel}$ modes (Figure~\ref{fig:pkslice}), considerably restricting the foreground leakage. Further, the smaller oscillation amplitude in the Cross \cl{} results in significantly lower power in the Cross $P(k_{\perp},k_{\parallel})$ beyond the theoretically predicted wedge boundary $[k_{\parallel}]_H$. 

The Cross $P(k_{\perp},k_{\parallel})$ heatmap (Figure~\ref{fig:pscyl}) shows that although the bulk of the foregrounds lie inside $[k_{\parallel}]_H$, there is a considerable amount of leakage beyond it. We have avoided these foreground-dominated modes (FG modes) and selected the relatively foreground free $21$-cm window (TW) region to put constraints (upper limits) on the cosmological $21$-cm signal. We have checked the noise statistics of the PS in the TW region through the quantity $X$ (defined in equation~\ref{eq:xstat}) which is expected to follow a standard normal distribution if the values of $P(k_{\perp},k_{\parallel})$ in the TW region are entirely due to the system noise. The PDF of $X$, in the central region $(\mid X \mid \leq 30 )$, is found to be mostly symmetric  with a positive mean $\mu = 0.61$ and a standard deviation $\sigma_{Est} = 4.77$ with no negative outlier values (Figure~\ref{fig:noisestat}). The absence of large negative outliers ensures that systematic, like discontinuities in the band edges, large phase errors, etc., are not affecting our power spectrum results. The standard deviation $\sigma_{Est} > 1$ suggests that the statistical fluctuations in $P(k_{\perp}, k_{\parallel})$ are underestimated by the system noise only simulations. To deal with this, we have scaled up our error estimates by the $\sigma_{Est}$ factor. On a related note, we find that (similar to \citetalias{P22}) a t-distribution adequately describes the central region of the PDF, but it fails to fit the positive tail. We show that a Lorentzian distribution represents the PDF better. We also notice the positive tail in the PDF to be more restricted (as compared to that of Total PS shown in \citetalias{P22}) in the  Cross PS analysis suggesting a cleaner (less foreground contaminated) TW region.

In Section~\ref{sec:BMLE}, we have explored the fact that the $21$-cm signal is isotropic in three-dimensional Fourier space, and hence its fluctuations can be entirely quantified with a spherical $21$-cm PS $P(k)$. We introduced a maximum likelihood estimator (MLE) which estimates $P(k)$ directly from the measured \cl{} without explicitly estimating $P(k_{\perp},k_{\parallel})$. We model the measured \cl{} as a combination of foregrounds, $21$-cm signal, and residual systematics (equation~\ref{eq:Cmodel}). We use the FG modes to model the foregrounds (equation~\ref{eq:CFG}), and the TW modes to model the $21$-cm signal (equation~\ref{eq:CTPK}). We incorporate the isotropy of the $21$-cm signal by dividing the TW region into spherical bins where the amplitude of the $21$-cm PS $P(k)$ is a constant. This approach is further illustrated with a schematic diagram in Figure~\ref{fig:diagram_mle}. We maximize the likelihood $\mathcal{L} \propto \exp{(-\chi^2 /2)}$, where the $\chi^2$ is defined through equation~(\ref{eq:chi2}), to find the best fit values of the free model parameters, $P(k)$ and the amplitudes of the FG modes $[P(k_{\perp},k_{\parallel})]_{FG}$. We have validated the MLE using simulations in Appendix~\ref{sec:validation}, and also presented an MCMC analysis (Appendix~\ref{sec:mcmc}) to show the consistency of the best-fit MLE solutions and their error estimates. The MCMC analysis also shows that the parameters $P(k)$ are uncorrelated in different spherical $k$-bins. The maximum likelihood estimation of the $21$-cm PS is likely to be more robust (than spherical averaging) in the presence of outliers \citep{Huber}. This framework is also more suitable for propagating any correlation between FG and $21$-cm modes. Thus, this method provides a self-consistent way of determining unbiased error bars on the $21$-cm power spectrum modes. We expect the error estimation will play an increasingly important role when IM experiments come close to making the first detections. 

We have used the best fit values of $P(k)$ to recover the isotropic component $[C_{\ell} (\Delta\nu)]_T$ which is found to be largely consistent with the noise fluctuations at $2\sigma$ level (Figure~\ref{fig:mle_PK}). The brightness temperature fluctuation, $\Delta^2(k)$, are also found to be consistent with the noise at $2\sigma$ level in most of the $k$ bins. The $\Delta^2(k)$ values are found to be $6-20$ times tighter than \citetalias{P22}, and are comparable to the findings of \citetalias{Ch21} (Figure~\ref{fig:pssph_8mhz}). The tightest constraint on the upper limits $\Delta_{UL}^2(k)\leq (58.67)^2 \, \rm{mK}^2$ at $k = 0.804 \, \rm{Mpc}^{-1}$ suggests $[\Omega_{\HI} b_{\HI}]_{UL} \leq 0.072$. These results are tabulated in Table~\ref{tab:ul_MLE}. The upper limits are nearly $5.2$ times better than our earlier results (\citetalias{P22}), where we have reported $\Delta_{UL}^2(k) \leq (133.97)^2 \, \rm{mK}^2$ and $[\Omega_{\HI} b_{\HI}]_{UL} \leq 0.23$ at $k = 1.03 \, \rm{Mpc}^{-1}$. Note, \citetalias{Ch21} results ($\Delta_{UL}^2(k) \leq (61.49)^2 \, \rm{mK}^2$ and $[\Omega_{\HI} b_{\HI}]_{UL} \leq 0.11$ at $k = 1 \, \rm{Mpc}^{-1}$) are similar to our current upper limits.

Finally, we have also considered the possibility of using the entire set of \cl{} measurements to directly constrain a single parameter $[\Omega_{\HI}b_{\HI}]^2$, without involving the intermediate step of estimating $P(k)$. In this approach, we model the $21$-cm signal using equation~\ref{eq:CTPK_Omb}, and estimate $[\Omega_{\HI} b_{\HI}]^2$ using the MLE. We find by combining four $\ell$-bins, as quoted in Table~\ref{tab:ul_omb} (Set II), $[\Omega_{\HI}b_{\HI}]^2 = 7.51\times 10^{-4} \pm 1.47\times 10^{-3}$ which is attributable to noise at $1\sigma$. Although the $2\sigma$ upper limit $[\Omega_{\HI}b_{\HI}]_{UL} \leq 0.061$ is  $\sim 50$ times larger than the expected value (see, e.g. \citealt{hamsa2015}), this is a considerable improvement over earlier IM works at this redshift. 
 
Although the upper limit is a significant improvement over \citetalias{P22}, a tighter constraint on the upper limit is expected if we can completely remove foregrounds from the data. Recently, \cite{Trott2022} have used a smooth foreground filter DAYENU \citep{Ewall-Wice2021} and estimated the  MAPS  \maps{} from a high-redshift $(z=6.2-7.5)$ Murchison Widefield Array (MWA; \citealt{Tingay2013}) data. Furthermore, the full MAPS  $C_{\ell}(\nu_a, \nu_b)$,  which does not assume the 21-cm signal to be ergodic  \citep{Mondal2018}, also provides possibilities for foreground removal  using eigendecomposition as presented in  \cite{liu12}, and also discussed in \cite{Mondal2022}. The idea is that the foregrounds, being featureless in frequency, can be accurately captured by means of only a few of the leading eigenmodes and this can be used to subtract out the foreground contribution. We plan to consider these possibilities in future work.

\section*{Acknowledgements}

We thank the anonymous reviewer for a careful reading of the manuscript and for the insightful comments and suggestions. We thank the staff of GMRT for making this observation possible. GMRT is run by National Centre for Radio Astrophysics of the Tata Institute of Fundamental Research. AE thanks Sukhdeep Singh for valuable discussions. AG would like to acknowledge IUCAA, Pune for providing support through the associateship programme. SB would like to acknowledge funding provided under the MATRICS grant SERB/F/9805/2019-2020 of the Science \& Engineering Research Board, a statutory body of Department of Science \& Technology (DST), Government of India. A part of this work has used the Supercomputing facility `PARAM Shakti' of IIT Kharagpur established under National Supercomputing Mission (NSM), Government of India and supported by Centre for Development of Advanced Computing (CDAC), Pune.

\section*{Data Availability}

The data used are available upon reasonable request to the corresponding authors.



\bibliographystyle{mnras}
\bibliography{myref} 



\appendix
\section{Validation of Cross estimator}
\label{sec:validation}

In \citetalias{P22} we have validated the Total correlation TGE (equation~\ref{eq:selfmaps}) using simulated visibilities corresponding to a sky signal which is assumed to be a Gaussian random field with a power spectrum $P^m(k)$
\begin{equation}
    \label{eq:modelps}
    P^{m}(\mathbf{k})  = A \left( \frac{k}{k_0}\right)^n {\rm  mK^{2} \,  Mpc^{3}}
\end{equation}
having an arbitrarily chosen value of $A=10$, $k_0 = 1 \, \mathrm{Mpc}^{-1}$, and a power law index $n=-2$. The simulated visibilities incorporated the same parameters (such as the baseline distribution, flagging etc.) of the data being used here. The details of the simulations can be found in \citetalias{P22}. The present work uses the Cross TGE (equation~\ref{eq:crossmaps}) which we validate here using the same simulated data.

\begin{figure}
	\includegraphics[width=\columnwidth]{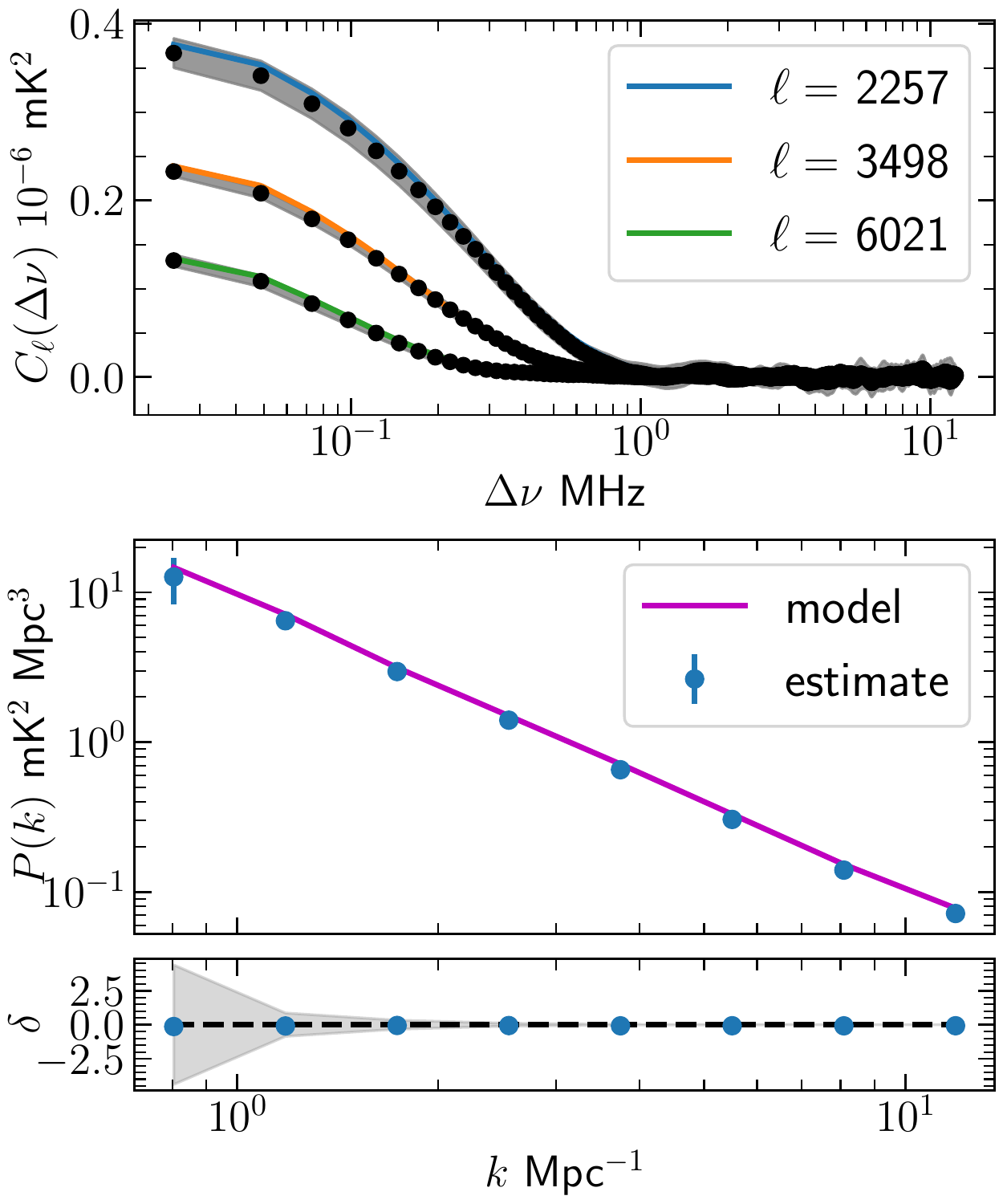}
    \caption{The uppermost panel shows the validation of Cross TGE (equation~\ref{eq:crossmaps}). The data points (black filled circles) show the mean $C_{\ell}(\Delta\nu)$ with $2\,\sigma$ errors (shaded region) estimated from 16 realizations of the simulated sky signal. The solid lines show the analytical predictions of $C_{\ell}(\Delta\nu)$ corresponding to the input model  $P^m(k)$.
    The middle panel shows the estimated spherically-binned power spectrum $P(k)$ (blue filled circles) and $2\,\sigma$ error bars estimated using the MLE. The input model $P^{m}(k)$ is shown with the purple solid line. 
    The bottom panel shows the fractional deviation  $\delta=[P(k)-P^m(k)]/P^m(k)$ (data points) and the expected $2 \sigma$ statistical fluctuations for  the same (grey shaded region).
    }
    \label{fig:validation}
\end{figure}

We have applied the Cross estimator (equation~\ref{eq:crossmaps}) on the simulated visibilities, and analyzed the simulated data identical to the actual data, to estimate $C_{\ell}(\Delta\nu)$.  We have used $16$  independent realizations of the simulation to estimate the  mean $C_{\ell}(\Delta\nu)$ and the $2\,\sigma$ errors shown in the uppermost panel of Figure~\ref{fig:validation} at three representative values of  $\ell$. We have also shown (solid lines) the analytical model predictions   calculated using  equation~\ref{eq:cl_Pk}. We see that the $C_{\ell}(\Delta\nu)$  estimated from the simulations closely matches the analytical predictions, which are mostly within the shaded region showing the $2\,\sigma$ uncertainty.

The last two panels show the validation of the MLE as a power spectrum estimator. The middle panel shows the estimated spherical PS $P(k)$ (blue filled circles) and $2\,\sigma$ error bars due to the cosmic variance. The input model $P^{m}(k)$ is shown with the magenta solid line.  We see that $P(k)$ is in reasonably good  agreement with $P^{m}(k)$ across the entire $k$ range considered here. The bottom panel shows the fractional deviation  $\delta=[P(k)-P^m(k)]/P^m(k)$ (data points) and the expected $2 \sigma$ statistical fluctuations for the same (grey shaded region). We have $\mid \delta \mid \,\, \lesssim  10 \%$ in most of the $k$-bins shown here. We see that the $\delta$ values are all consistent with the predicted $2 \sigma$ errors. We have somewhat larger error bars at the smallest $k$-bin. The convolution with the tapering window function (equation~\ref{eq:vcgx}) is expected to become important at the small baselines \citep{samir14}, and this possibly contributes to enhancing the statistical fluctuations in the small $k$-bins. A part of the deviations could also arise from the low baseline densities in some of the bins. The entire validation presented here used {\it exactly}  the same $(k_{\perp},k_{\parallel})$ modes as those that have been used for the actual data. In summary, we have validated the Cross estimator and find that it can recover the input model PS to an accuracy close to $\lesssim 10\%$ across the entire $k$ range considered here.

\section{MCMC Analysis}
\label{sec:mcmc}

A Markov Chain Monte Carlo (MCMC) algorithm allows us to draw parameter samples that are consistent with the measured data. Using the parameter samples, we construct the probability distribution of the model parameters and also characterize the correlation between the parameters. Here, we use an MCMC to check the consistency of the best fit values and their error estimates which we have obtained using the maximum likelihood estimator presented in Section~\ref{sec:BMLE} and \ref{sec:omb}. For this purpose, we first consider the posterior probability distribution of the parameters $[P(k)]_T$. We keep $[P(k_{\perp}, k_{\parallel})]_{{\rm FG}}$, the amplitudes of the FG modes, which are the other free parameters of our model, fixed at the maximum likelihood estimates. Since the FG and the TW modes are uncorrelated (assuming the initial density fluctuations are Gaussian on large scales), keeping $[P(k_{\perp}, k_{\parallel})]_{{\rm FG}}$ fixed does not change the posterior of $[P(k)]_T$. We have used a uniform prior $\mathcal{U}(-\infty, \infty)$ on $[P(k)]_T$, this allows $[P(k)]_T$ to have any possible real numbers with an equal probability. The prior, along with the likelihood defined through equation~\ref{eq:chi2}, yields the posterior ($\propto$ likelihood $\times$ prior) from which we draw samples using an MCMC.

\begin{figure*}
    \centering
	\includegraphics[width=\textwidth]{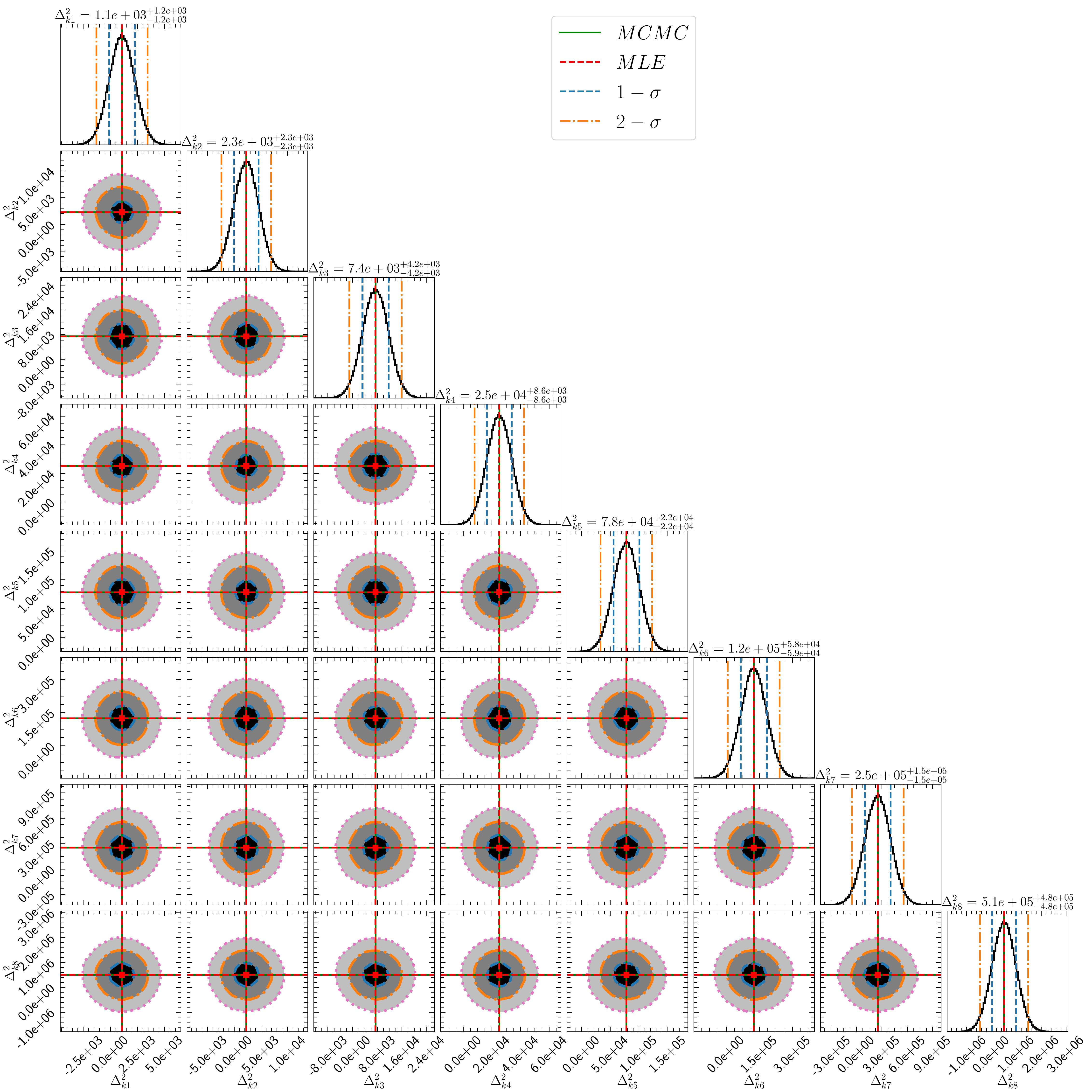}
    \caption{The diagonal shows the marginalized one-dimensional posterior probability distribution of $\Delta^2(k)$ in ${\rm mK^2}$. The vertical lines display the mean (solid green) and the associated $1\sigma$ (dashed blue) and $2\sigma$ (dashed-dot orange) standard deviations of the  MCMC samples along with the best-fit value from MLE (red dashed).
    Each panel of the off-diagonal show the marginalized two-dimensional projections of the posterior probability distribution of each pair of the parameters. The  blue (dashed),  orange (dashed-dot) and  pink (dotted) contours show the $1\sigma$, $2\sigma$, and $3\sigma$ levels, respectively. The  green (solid) lines and the  red (dashed) lines show the MCMC median and the MLE solutions respectively. 
    The plot has made use of the  python module \textsc{corner} \citep{corner}.}
    \label{fig:corner}
\end{figure*}

We have used the affine-invariant ensemble sampling (AIES; \citealt{aies}) algorithm implemented in the python module {\scriptsize EMCEE} \citep{mcmc} to draw samples from the posterior probability distribution of $[P(k)]_T$. 
Figure~\ref{fig:corner} shows the posterior probability distributions of $\Delta^2(k)$ which is obtained by scaling the MCMC samples of $[P(k)]_T$ with $k^3/2\pi^2$. The panels in the main diagonal show the one-dimensional marginalized posterior probability distribution of $\Delta^2(k)$, whereas, the off-diagonal panels show the two-dimensional projections of the posterior probability distribution of each pair of the parameters. In each panel, the solid green lines show the mean value of $\Delta^2(k)$ derived from the MCMC samples. The maximum likelihood solutions (hereafter, MLE solutions) which are obtained from maximizing the likelihood (Section~\ref{sec:BMLE}),  are also shown (dashed red lines) along with the MCMC solutions. Note, all values quoted in the figure are in $\mathrm{mK^2}$ units. Considering the diagonal panels, the dashed blue and the dashed-dot orange vertical lines demarcate the $1\sigma$ and $2\sigma$ levels respectively. In the off-diagonal panels, the  blue (dashed),  orange (dashed-dot) and  pink (dotted) contours show the $1\sigma$, $2\sigma$ and $3\sigma$ levels respectively. For all the $k$ values considered here, the MLE solutions are found to lie within the $1\sigma$ uncertainty intervals of the MCMC solutions. We also do not find any correlation among the parameters. The error estimates are also quite similar in both analyses. We have also computed the $2\sigma$ upper limits on $\Delta^2(k)$ and $[\Omega_{\HI}b_{\HI}]$ using the MCMC samples. The MCMC means, their uncertainties and the upper limits are highlighted in Table~\ref{tab:ul_MCMC}.  
\begin{table}
    \centering
    \caption{Same as Table~\ref{tab:ul_MLE} except that the mean value of $\Delta^2(k)$ and standard deviations $\sigma$ are estimated using an MCMC. }
        \begin{tabular}{cccccc}
        \hline
        \hline
        $k$ & $\Delta^2(k)$ & $1\sigma$ & SNR & $\Delta_{UL}^{2}(k)$ & $[\Omega_{\HI}b_{\HI}]_{UL}$ \\
        Mpc$^{-1}$ & (mK)$^2$ & (mK)$^2$ & & (mK)$^2$  &\\
        \hline
        $0.804$ &  $(32.71)^2$ & $(34.33)^2$ & $0.908$ & $(58.55)^2$ & $0.072$\\
        $1.181$ &  $(47.60)^2$ & $(48.17)^2$ & $0.976$ & $(83.11)^2$ & $0.089$\\
        $1.736$ &  $(86.16)^2$ & $(65.19)^2$ & $1.747$ & $(126.19)^2$ & $0.121$\\
        $2.551$ &  $(158.63)^2$ & $(93.12)^2$ & $2.902$ & $(206.17)^2$ & $0.177$\\
        $3.748$ &  $(279.53)^2$ & $(149.72)^2$ & $3.486$ & $(350.67)^2$ & $0.273$\\
        $5.507$ &  $(353.38)^2$ & $(242.35)^2$ & $2.126$ & $(492.29)^2$ & $0.350$\\
        $8.093$ &  $(501.87)^2$ & $(390.10)^2$ & $1.655$ & $(745.81)^2$ & $0.488$\\
        $11.892$ &  $(711.87)^2$ & $(696.66)^2$ & $1.044$ & $(1215.50)^2$ & $0.588$\\
        \hline
    \end{tabular}
    \label{tab:ul_MCMC}
\end{table}

\begin{figure}
	\includegraphics[width=\columnwidth]{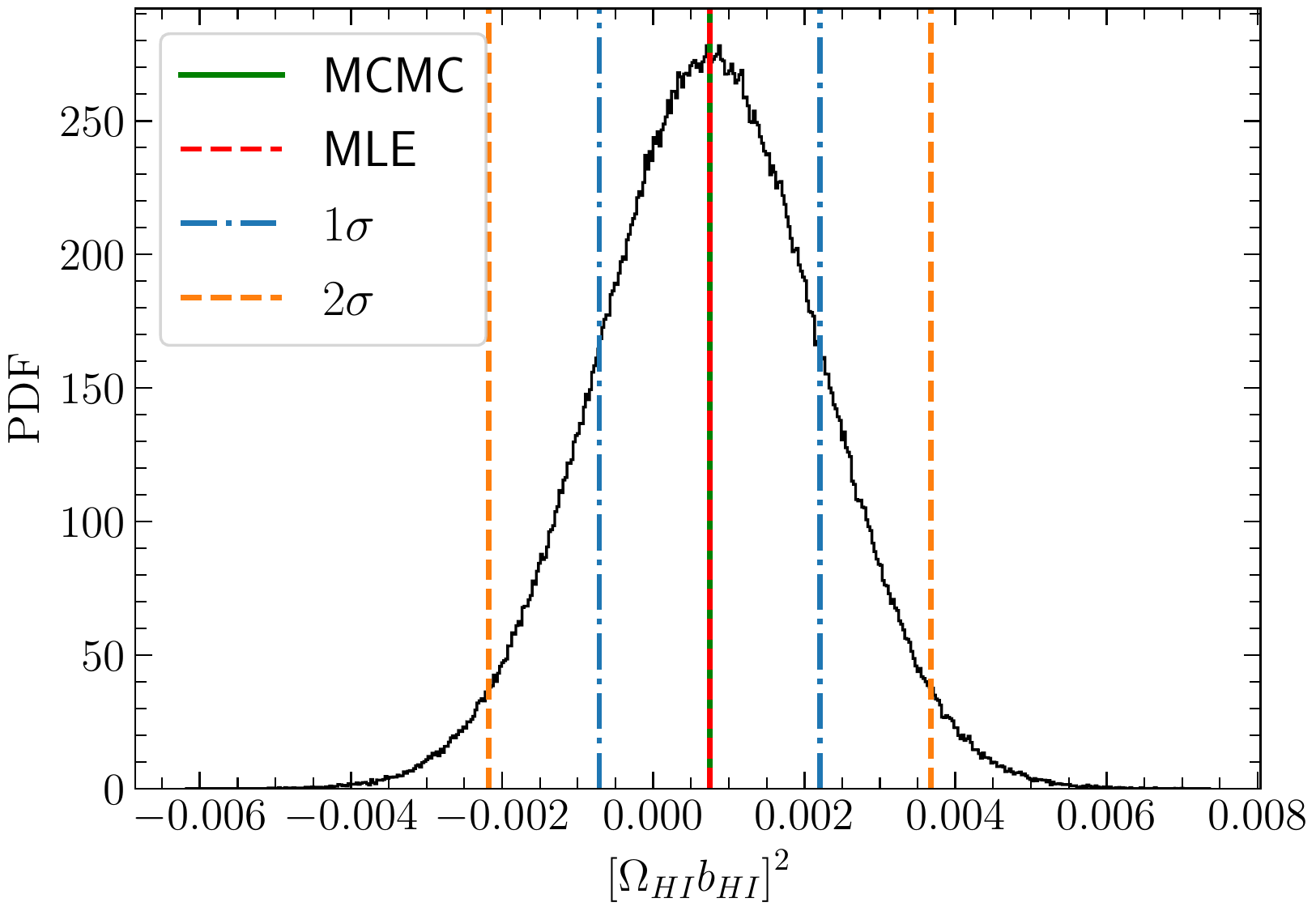}
    \caption{This figure shows the posterior probability distribution of $\left[\Omega_{\HI} b_{\HI}\right]^2$. The vertical lines show the mean (solid green), the associated $1\sigma$ (dashed blue) and $2\sigma$ (dashed-dot orange) errors estimated from the MCMC samples, along with the best-fit value obtained from the MLE (red dashed) in Section~\ref{sec:omb}.} 
    \label{fig:posterior_omb}
\end{figure}

Next, we carry out an MCMC analysis to constrain $[\Omega_{\HI} b_{\HI}]$, which we have previously done using the MLE in Section~\ref{sec:omb}. We have used the measured \cl{} values from the same $4$ $\ell$ bins mentioned in the Set II of Table~\ref{tab:ul_omb}. We have kept the $[P(k_{\perp}, k_{\parallel})]_{{\rm FG}}$ values fixed at the maximum likelihood estimates, and used a uniform prior $\mathcal{U}(-2, 2)$ on $[\Omega_{\HI} b_{\HI}]^2$. We have also checked that a broader range on the prior does not alter the posterior. Figure~\ref{fig:posterior_omb} shows the resulting posterior probability distribution of the model parameter $\left[\Omega_{\HI} b_{\HI}\right]^2$. The vertical lines show the mean (solid green) and the associated $1\sigma$ (dashed blue) and $2\sigma$ (dashed-dot orange) uncertainties along with the best-fit value obtained from MLE (red dashed). We find $[\Omega_{\HI} b_{\HI}]^2 = 7.50\times10^{-4}\pm1.46\times10^{-3}$ which translates into a $2\sigma$ upper limit  $[\Omega_{\HI} b_{\HI}]_{UL} \leq  6.06\times10^{-2}$. These values are found to be close to the MLE solutions. 

In both the MCMC runs, we have used $200$ random walkers and initialized them to the vicinity of the parameters derived from the MLE to ensure a faster convergence. We first performed $300$ burn-in steps before running the full chain of $10000$ steps. To check the convergence of the MCMC, we have considered the quantity $\tau_f$, the integrated autocorrelation time, which gives an estimate of the number of steps required for the chains to converge \citep{aies}. Here we quote the mean autocorrelation time, which is the mean of the integrated autocorrelation time estimated for the chains corresponding to each parameter. The mean autocorrelation time is found to be $\sim82$ and $\sim26$ steps for the $8$ and $1$ parameter cases of $P(k)$ and  $[\Omega_{\HI} b_{\HI}]^2$, respectively. We have conservatively chosen a significantly large number of steps $(>50 \times \tau_f)$ to reduce the sample variance. We have also checked the trace plots of the MCMC to ensure the convergence of AIES.

\section{A Comparison}
\label{sec:comparison}

\begin{figure}
	\includegraphics[width=\columnwidth]{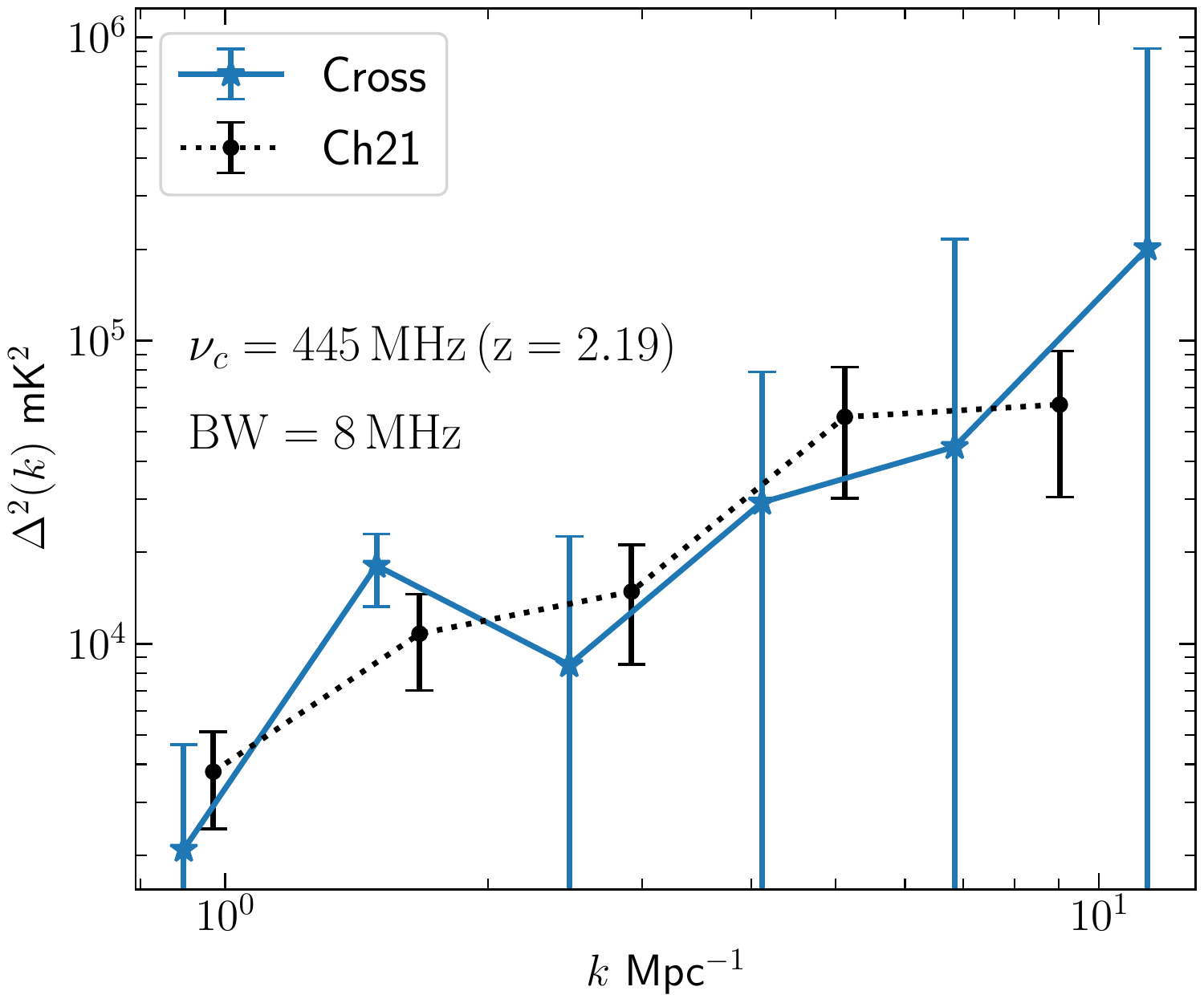}
    \caption{The mean squared brightness temperature fluctuations $\Delta^2(k)$ along with $2 \sigma$ error bars from the analysis of an $8$ MHz bandwidth data at $\nu_c=445$ MHz $(z=2.19)$. The blue solid and the black dotted lines show the $\Delta^2(k)$ values obtained from  the present work and \citetalias{Ch21}, respectively.}
    \label{fig:pssph_8mhz}
\end{figure}

In this appendix we present an analysis of the same $8$ MHz bandwidth data as analysed by \citetalias{Ch21} for a comparison between the two different estimators used in these two works. The data is drawn from the central frequency $\nu_c = 445 \,\rm{MHz}$ which corresponds to the redshifted $21$-cm signal from $z=2.19$. In short, we have used the Cross TGE (equation~\ref{eq:crossmaps}) to estimate \cl{}, and use the MLE (Section~\ref{sec:BMLE}) to estimate the spherical PS $P(k)$ at several $k$-bins. The blue solid line in Figure~\ref{fig:pssph_8mhz} shows the $\Delta^{2}(k)$ values obtained from the present analysis, whereas the black dotted line shows the same as shown in the middle panel of the Figure~A1 of \citetalias{Ch21}. We find that the $\Delta^{2}(k)$ values in the two analyses are comparable and consistent within the $2\sigma$ error bars. The error bars in the present analysis is found to be larger $(\sim\sqrt{2}\,\rm{times})$ than \citetalias{Ch21}. This larger error bars in the present analysis can be attributed to the fact that we have used only the correlation between the cross-polarizations (RR $\times$ LL) and discarded the correlation of the self-polarizations (RR $\times$ RR and LL $\times$ LL). For comparison, \citetalias{Ch21} reported $2\sigma$ upper limits $\Delta_{UL}^2(k)\leq (61.49)^2 \, \rm{mK}^2$ at $k = 0.97 \, \rm{Mpc}^{-1}$, whereas we find $\Delta_{UL}^2(k)\leq (68.13)^2 \, \rm{mK}^2$ at $k = 0.90 \, \rm{Mpc}^{-1}$.

A notable feature in TGE is that by tapering, it restricts the wide-angle point source contributions to comparably small $k_{\parallel}$ and thereby broadens the accessible TW region, which enables us to probe larger scales. We also found that (Figure~\ref{fig:pkslice}) the Cross TGE significantly reduces various polarization-dependent systematics in \cl{}, allowing access to the smaller $k_{\parallel}$ modes. As the two estimators yield different (relatively) foreground-free TW regions, we have not attempted to compare them in a one-to-one basis (i.e. same $k$-value).

In addition to the cross-polarization PS described here, another significant distinction between the TGE and the techniques employed \citetalias{Ch21} is the treatment of the missing frequency channels which are flagged due to RFI. This issue is highlighted in \citetalias{P22} (Section 5), which we briefly reiterate here. For each baseline, \citetalias{Ch21} have computed a Fourier transform of the measured visibilities along frequency to estimate the delay space visibilities \citep{Morales04}, which are then used \citep{parsons12} to estimate the PS. The missing frequency channels introduce ringing artefacts in the Fourier transform and corrupt the estimated PS. \citetalias{Ch21} have used the 1D CLEAN introduced by \cite{Parsons_2009} to get uncorrupted delay space visibilities from RFI-contaminated data. This 1D CLEAN, which is adapted \citep{Roberts1987} from the two-dimensional \textsc{CLEAN} deconvolution algorithm \citep{Hogbom1974} used in aperture synthesis, performs a nonlinear deconvolution in the delay space, equivalent to a least-squares interpolation in the frequency domain. In contrast, the TGE first correlates the visibility data across frequency channels to estimate $C_{\ell}(\Delta \nu)$. Despite having a substantial number of  missing frequency channels in the visibility data ($55\%$ here), there are no missing  frequency separations $\Delta \nu$ in the estimated  $C_{\ell}(\Delta \nu)$. The MLE (Section~\ref{sec:BMLE}) is then used to estimate the spherical PS $P(k)$ from the \cl{}. It is not essential to make up for any missing frequency channels because the entire procedure uses only the available frequency channels to estimate the PS. \cite{Bh18} has used simulations to demonstrate that TGE can successfully recover the PS even  when the data in $80 \%$ randomly chosen frequency channels are flagged. For the present analysis, we have validated the estimator (Appendix~\ref{sec:validation}) using simulations where the flagging of the simulated data exactly matches that of the actual data. In addition to this direct validation of the estimator using simulated data, this comparison with \citetalias{Ch21}, and the broadly consistent match of $\Delta^{2}(k)$ values from the two very distinct methods make our results assuredly more reliable.

\bsp	
\label{lastpage}
\end{document}